\documentstyle[epsfig,here,12pt]{JHEP}

\title{Conformal window and Landau singularities}
\author{G.\ Grunberg\thanks{Research
supported in part by the EC program ``Training and 
Mobility of 
Researchers'', Network ``QCD and Particle Structure'', contract 
ERBFMRXCT980194.}\\
        Centre de Physique Th\'eorique de l' Ecole  
Polytechnique (CNRS UMR C7644),\\
        91128 Palaiseau Cedex, France\\
        E-mail: \email{grunberg@cpht.polytechnique.fr}}
\abstract{A physical characterization of Landau singularities is emphasized, which should trace the lower boundary
$N_f^*$ of the conformal window in QCD and supersymmetric QCD. A natural way to disentangle ``perturbative'' from
 ``non-perturbative'' contributions 
 below $N_f^*$ is suggested. Assuming an infrared fixed point is present in the perturbative part of the
QCD coupling even in some range below $N_f^*$ leads to the condition $\gamma(N_f^*)=1$, where $\gamma$ is the critical
 exponent. This result is incompatible with the existence of an analogue of Seiberg duality
 in QCD. Using the Banks-Zaks expansion, one gets 
 $4\leq N_f^*\leq 6$. The low value of $N_f^*$ gives some justification to the  infrared finite coupling
approach to power corrections, and suggests a way to compute their normalization from perturbative
input. If the perturbative series are still asymptotic in the negative coupling region, 
the presence of a negative ultraviolet fixed point is required both in QCD and in supersymmetric QCD
 to preserve causality 
within the conformal window. Some  evidence for such a fixed point in QCD is provided through a modified Banks-Zaks expansion.
Conformal window amplitudes, which contain power contributions, are shown to remain generically finite in the
$N_f\rightarrow -\infty$ one-loop limit in simple models with infrared finite perturbative coupling.}
\preprint{CPTh/S 016.0401}
\begin{document}
\section{Introduction}
The notion of an infrared (IR) finite  coupling has been used \cite{Dok} extensively in recent years, especially
in connection with the phenomenology of power corrections in QCD. The present investigation  is motivated by the
desire to understand better the theoretical background behind such an assumption. In particular, given
an IR finite coupling $\bar{\alpha}$, does it remain  finite within perturbation
theory itself (such as the two-loop coupling  with opposite signs one and two loop beta function coefficients), or does one need a
non-perturbative contribution $\delta\alpha$ to cancel ($\bar{\alpha}=\alpha+\delta\alpha$) the Landau singularities present in its
  perturbative part $\alpha$? The answer I shall suggest is a mixed one: the perturbative
part of the
QCD coupling may be {\em always} IR finite but, below the so called ``conformal window'' (the range of $N_f$ values where the theory
is scale invariant at large distances and flows to a non-trivial IR fixed point), 
one still needs a $\delta\alpha$
term since the perturbative coupling is no more causal there, despite being IR finite. As the main outcome, one obtains
 an equation to determine
the lower boundary $N_f^*$ of the conformal window in QCD. One finds a low value of $N_f^*$, which, as we 
shall see, gives some justification to the infrared finite coupling approach to power corrections.
 The plan of the paper is as follows. In section 2 I review
 the evidence
and present a formal argument
for the existence of Landau singularities in the perturbative coupling. A more physical argument, relating Landau
singularities to the very existence of the conformal window and a two-phase structure of QCD is given in section 3, which
also suggests a clean way to disentangle ``perturbative'' from ``non-perturbative'' contributions below the conformal
 window. In section
4, two main scenarios for causality breaking are described. In section 5, an equation to determine the bottom of the conformal
window in QCD is suggested, and is solved through the Banks-Zaks expansion in section 6. 
Section 7 gives evidence, through
a modified Banks-Zaks expansion, for the existence of a {\em negative} ultraviolet (UV) fixed point in QCD, necessary for the
consistency of the present approach. In section 8, the disentangling between``perturbative'' and
 ``non-perturbative'' components of condensates below the conformal window is performed explicitly in the so-called ``APT'' model
for the non-perturbative coupling. Section 9 presents a  justification to the IR finite coupling approach to
 power corrections, and suggests a possibility to actually {\em compute} the main contribution to their 
normalization from perturbative input. The issue
whether conformal window amplitudes (which include power terms) remain finite in the $N_f\rightarrow -\infty$ limit
as suggested by the behavior of the corresponding perturbative series is discussed in section 10.
Section 11 contains the conclusions. More technical details are dealt with in two appendices. Appendix A gives the proof
of a necessary condition for causality. A modified Banks-Zaks expansion is developed in Appendix B.
Appendix C derives the form of power corrections in models
 with non-trivial IR fixed points. A shorter version of some of the present results appeared in \cite{Gru-short}.

\section{Evidence for Landau singularities in the perturbative coupling}

The only present evidence for a Landau singularity in the  perturbative 
{\em renormalized}\footnote{In QED, the well established ``triviality'' property gives only direct
 evidence \cite{G-qed}
for a singularity in the {\em bare} coupling constant.}
 coupling is still
the old Landau-Pomeranchuk leading log QED calculation, now reformulated in QCD as a $N_f\rightarrow -\infty$
(``large $\beta_0$'') limit.  In this limit, the perturbative coupling is one-loop
\begin{equation}\alpha(k^2)={1\over \beta_0 \log{k^2\over \Lambda^2}}\label{eq:1-loop}
\end{equation}
where $\Lambda$ is the Landau  pole. The question is whether there is a singularity
at {\em finite} $N_f$. Some light on this problem can be shed by considering further the $N_f$ dependence. Indeed,
 another (conflicting) piece of information is available at the other end of the spectrum,
 around the value $N_f=N_f^0=16.5$ (I consider $N_c=3$) where the one loop
 coefficient $\beta_0={1\over 4}(11-{2\over 3}N_f)$
of the beta function vanishes (``small $\beta_0$'' limit). For $N_f$ slightly below $16.5$ 
 a weak coupling (Banks-Zaks) 
IR fixed point develops \cite{BZ,White,G-bz}, and the perturbative coupling is causal beyond one-loop, 
 i.e. there are no Landau singularities in 
the whole first sheet of the complex $k^2$ plane. Can then the perturbative coupling remain causal down to 
$N_f=-\infty$? I shall assume that the limit of a sequence of causal couplings must itself be causal. Indeed, a causal coupling satisfies the dispersion
relation

\begin{equation}\alpha(k^2)
=  -\int_0^\infty{d\mu^2\over
\mu^2+k^2}\
\rho(\mu^2)\label{eq:PT-causal}\end{equation}
and the previous statement follows if one can take the limit under the integral.
In such a case, the existence of 
a Landau pole
at $N_f\rightarrow -\infty$ implies the existence of a {\em finite} value $N_f^*$  below which 
Landau singularities appear on the first sheet of the complex $k^2$ plane and perturbative causality
is lost, which is the common wisdom (at $N_f^*$ itself, according to the above philosophy, the 
coupling must still be causal). The range $N_f^*<N_f<N_f^0$ where the {\em perturbative} coupling is causal
and flows to a finite IR fixed point
is taken as the definition of  the ``conformal window'' for the sake of the present discussion
(this definition will be refined in the next section).
 I shall propose in section 5 an ansatz to determine $N_f^*$ (the bottom of the conformal window) in QCD,
 but first I give a more physical argument in favor of the existence of Landau singularities, which also illuminates
their physical meaning.

\section{Landau singularities and conformal window}
Let us assume the existence of a two-phase structure in QCD as the number of flavors $N_f$ is varied: 

i) For $N_f^*<N_f<N_f^0$  (the conformal window) the theory is scale invariant at large distances, and the vacuum is 
``perturbative'', in the sense there is no confinement nor chiral symmetry breaking. Conformal window amplitudes (generically
noted as $D_{\overline{PT}}(Q^2)$, where $Q$ stands for an external scale) are in this generalized sense ``perturbative'',
 i.e. could in principle be determined from information
contained in perturbation theory to all orders. 
Note that, even barring instantons contributions, $D_{\overline{PT}}(Q^2)$ is expected to include power corrections
terms (the so-called``condensates'', see section 8) which are usually viewed as typically non-perturbative:
 this motivates the subscript $\overline{PT}$.

ii) For $0<N_f<N_f^*$ there is a phase transition to a non-trivial vacuum, with confinement and chiral symmetry breaking, as expected in 
standard QCD.

A direct, {\em physical} motivation for Landau singularities can now be given: they trace the lower boundary $N_f=N_f^*$
of the conformal window. This statement is implied from the following two postulates:

1) Conformal window amplitudes $D_{\overline{PT}}(Q^2)$ can be analytically continued in $N_f$ below the bottom  $N_f^*$
of the conformal window.

2) For $N_f<N_f^*$, the (analytically continued) conformal window amplitudes $D_{\overline{PT}}(Q^2)$ must
 {\em differ} from the
 {\em full}  QCD amplitude $D(Q^2)$, since one enters
a new phase, i.e. we have

\begin{equation}D(Q^2)=D_{\overline{PT}}(Q^2)+D_{\overline{NP}}(Q^2)\label{eq:PT-NP}
\end{equation}
(whereas $D(Q^2)\equiv D_{\overline{PT}}(Q^2)$ within the conformal window).
 Assuming QCD to be a {\em unique} theory at given $N_f$, $D_{\overline{PT}}(Q^2)$ cannot provide a consistent
solution if $N_f<N_f^*$: this must be signalled by the appearance of unphysical Landau singularities
in $D_{\overline{PT}}(Q^2)$. $N_f^*$ should thus coincide with the value of $N_f$ below which (first sheet)
 Landau singularities  appear in
$D_{\overline{PT}}(Q^2)$. The occurrence of a ``genuine'' non-perturbative component $D_{\overline{NP}}(Q^2)$  is then 
necessary below $N_f^*$ in order to cancel the Landau singularities present in $D_{\overline{PT}}(Q^2)$.
 If these assumptions are correct, they provide an interesting connection between information contained
in principle in ``perturbation theory'' (including eventually all instanton sectors),
 which fix the structure of the conformal window
 amplitudes and ``genuine'' non-perturbative
phenomena, which fix the bottom of the conformal window. In addition, eq.(\ref{eq:PT-NP}) provide a neat way to disentangle
 the ``perturbative'' from the genuine ``non-perturbative'' part of an amplitude, for instance the part of the gluon
 condensate related to renormalons from the one reflecting
the presence of the non-trivial vacuum. Note also $D_{\overline{PT}}(Q^2)$ and $D_{\overline{NP}}(Q^2)$ are separately
free of renormalons ambiguities, but contain Landau singularities below $N_f^*$, so the renormalon and Landau
singularity problems are also disentangled (an example shall be provided in section 8).
In order to get a precise condition to determine $N_f^*$, we need now to look
in more details how causality can be broken in the perturbative coupling.

\section{Scenarios for causality breaking}

One can distinguish two main  scenarios:

i) The ``standard'' one where the IR fixed point $\alpha_{IR}$ present within the conformal window just disappears when
$N_f<N_f^*$ while a real, space-like Landau singularity is generated in the perturbative coupling. For instance
two simple zeroes of the beta function can merge into a  double zero when $N_f=N_f^*$  before moving to the complex plane.
 An example   is  afforded by 
the 3-loop beta function
\begin{equation}\beta(\alpha)={d\alpha\over d\log k^2}=-\beta_0 \alpha^2-\beta_1 \alpha^3-\beta_2 \alpha^4
\label{eq:3-loop}
\end{equation}
 with $\beta_0>0$, $\beta_2>0$ but $\beta_1<0$, in order to have a positive UV fixed point $\alpha_{UV}>\alpha_{IR}$.
 Starting from a situation where the IR fixed point is present (which
requires $\beta_1^2>4\beta_0\beta_2$) and the coupling is causal (which requires 
in addition \cite{Gar-Gru-conformal} that the 2-loop condition $0<-\beta_0^2/\beta_1<1$ be satisfied, see  section 7 for the
general explanation of this fact),
 one can imagine decreasing
$\vert\beta_1\vert$ (keeping $\beta_1<0$) down to the point $\beta_1^2=4\beta_0\beta_2$
 where the IR and UV fixed points coalesce
 before becoming
 complex
 and the physical IR fixed point disappears, which gives the bottom of the conformal window
in this model. Up to this point, the coupling
can still be causal if  $4\beta_2\geq \beta_0^3$.

Another possibility is that the zero $\alpha^*$ of the beta function is shielded by a pole $\alpha_P$, i.e. decreasing
 $N_f$ below the conformal
 window
one moves from a situation where $0<\alpha^*=\alpha_{IR}<\alpha_P$ to one where $0<\alpha_P<\alpha^*$. If the
 zero if of 
sufficiently high order that the beta function still vanishes in the limit where $\alpha^*=\alpha_P$ (which requires
at least a double zero in presence of a simple pole), this limit gives the bottom of the conformal window 
(provided the coupling is causal for  $0<\alpha_{IR}<\alpha_P$). Such a scenario is a plausible one in SQCD \cite{Seiberg,
Anselmi}.

It is also possible that the IR fixed point disappears by moving to infinity at the bottom of the conformal window
 (this mechanism does not require any extra zero). An example is provided by the ``Pad\'e-improved'' 3-loop
beta function \cite{Gar-Gru-Kar,Gar-Gru-conformal}

\begin{equation}\beta(\alpha)=-\alpha^2\ {\beta_0+(\omega\beta_0+\beta_1)\alpha\over 1+\omega\alpha}
\label{eq:3-loop-Pade}\end{equation}
with $\omega>0$ (hence the beta function pole is at negative coupling) and $\beta_1<0$. Then $\alpha_{IR}=-{\beta_0\over
\omega\beta_0+\beta_1}$
 diverges for $\beta_0=-\beta_1/\omega$,
 which gives the bottom of the conformal window (provided the coupling is causal for $0<\beta_0<-\beta_1/\omega$, which
 requires again \cite{Gar-Gru-conformal} that the 2-loop condition $0<-\beta_0^2/\beta_1<1$ be satisfied: this can be achieved
if $0<-\beta_1<\omega^2$).

ii) Alternatively, it is possible for the IR fixed point {\em to be still present}\footnote{This assumption is consistent
with the suggestion  \cite{Stev} that the {\em perturbative} coupling has a non-trivial IR fixed point down
 to $N_f=2$ in QCD.
 However
the full non-perturbative coupling must still differ by a $\delta\alpha$ term, since the perturbative coupling is
non-causal below $N_f=N_f^*$.}
 in the {\em perturbative part} of the coupling at least in some range
 $N_f^c<N_f<N_f^*$ {\em below} the conformal window. The motivation behind this assumption is the observation \cite{Stev,Gar-Kar,Gar-Gru-conformal} that, 
 for QCD effective
charges associated to Euclidean correlators (for which the notion of $k^2$ plane analyticity makes sense), 
the Banks-Zaks expansion in QCD (as opposed to SQCD \cite{Gar-Gru-conformal}) seems to converge down to fairly small values of $N_f$.
In this case
there can be no space-like Landau singularity, and 
causality must be violated by the appearance of
{\em complex} Landau singularities  on the first sheet of the  $k^2$ plane. It is natural to  assume,
as suggested by the 2-loop example below, 
that they arise 
as the result of the continuous migration  to the first sheet, through
 the time-like cut, of some second sheet  singularities already
present when $N_f>N_f^*$. I shall assume that this is the scenario which prevails in  QCD. As the simplest example, 
consider the two-loop coupling

\begin{equation}{d\alpha\over d\log k^2}=-\beta_0 \alpha^2-\beta_1 \alpha^3\label{eq:2-loop}
\end{equation}
If $\beta_0 >0$ but $\beta_1 <0$, there is an IR fixed point at $\alpha_{IR}=-{\beta_0\over \beta_1}$.
 It has been shown  \cite{Uraltsev,Gar-Gru-Kar,Gar-Gru-conformal} that this coupling has a pair of complex conjugate Landau singularities 
  on the second (or higher)
sheet if
 \begin{equation}0<\gamma_{2-loop}=-{\beta_0^2\over \beta_1}<1\label{eq:gamma-2loop}
\end{equation}
where $\gamma_{2-loop}$ is the (2-loop) critical exponent, the
 derivative of the beta function at the fixed point

\begin{equation}\gamma={d\beta(\alpha)\over d\alpha}\vert_{\alpha=\alpha_{IR}}
\label{eq:exponent}\end{equation}
(a simple proof in a more general case is given at the end of this section). For $\gamma_{2-loop}>1$, the second sheet singularities move to the first sheet through the time-like cut,
 which is reached when $\gamma_{2-loop}=1$. The latter condition thus determines the bottom of the 
conformal window in this model. Note that in the limit $\beta_1\rightarrow 0^-$ where $\gamma_{2-loop}=+\infty$,
 one gets the one loop coupling and the complex conjugate singularities collapse to  a  space-like Landau pole. This 
limit is thus the analogue of the $N_f\rightarrow -\infty$ limit in  QCD.

A somewhat more generic example (see section 7)
 is provided by the  3-loop beta function eq.(\ref{eq:3-loop}), this time
 with $\beta_0>0, \beta_2<0$
($\beta_1$ can have any sign) such that there is a positive IR fixed point, but a {\em negative} UV 
fixed point.  Causality is  obtained  for \cite{Gar-Gru-conformal}

\begin{equation}0<\gamma_{3-loop}<1\label{eq:gamma-3loop}
\end{equation}
where $\gamma_{3-loop}$ is the 3-loop critical exponent at the IR fixed point, and the bottom of the 
conformal window corresponds to $\gamma_{3-loop}=1$.

In the previous examples, the IR fixed point approaches $+\infty$ in the  one-loop limit where $\beta_i\rightarrow 0$ ($i\geq 1)$. 
It is possible however the  IR fixed point remains {\em finite}. 
 An
example is provided by a beta function with one positive pole $\alpha_P$ (the required Landau singularity) and two 
opposite sign
 zeroes (an IR and an UV fixed point) $\alpha_{IR}$ and $\alpha_{UV}$:
\begin{equation}\beta(\alpha)=-\beta_0\alpha^2\ {(1-\alpha/ \alpha_{IR}) (1-\alpha/ \alpha_{UV})\over
 1-\alpha/ \alpha_P}\label{eq:UV-IR-pole}\end{equation}
where $\alpha_{UV}<0$ and $0<\alpha_{IR}<\alpha_P$. The one-loop limit is achieved for $\alpha_{IR}=\alpha_P$
and $\alpha_{UV}=-\infty$. Although the IR fixed point remains finite, the corresponding critical exponent 
still\footnote{In this sense, this example is the opposite of the last one in point i) above.}
 approaches $+\infty$
 (as in the other examples), and one can check that
 causality is violated when it passes through $1$ (this statement
 also follows, in the particular case $\alpha_{UV}=\infty$ where one recovers the ``Pad\'e-improved'' 3-loop
beta function, from the results of \cite{Gar-Gru-Kar,Gar-Gru-conformal}).
Indeed, the solution of the corresponding renormalization group equation is

\begin{equation}\log{k^2\over\Lambda^2}={1\over\beta_0\alpha}+{1\over\gamma_{IR}}
\log\left({1\over\alpha}-{1\over\alpha_{IR}}\right)+{1\over\gamma_{UV}}
\log\left({1\over\alpha}-{1\over\alpha_{UV}}\right)\label{eq:RG-4loop-Pade}\end{equation}
where $\gamma_{IR,UV}$ are the critical exponents at the IR (UV) fixed points with

\begin{equation}\gamma_{IR}=\beta_0\ \alpha_{IR} {1-{\alpha_{IR}\over\alpha_{UV}}\over 1-{\alpha_{IR}\over\alpha_P}}
\label{eq:gamma-4loop-Pade}\end{equation}
and similarly for $\gamma_{UV}$ (with $IR\leftrightarrow UV$). The location (in $\alpha$ space) of the Landau singularities 
read directly from the beta function eq.(\ref{eq:UV-IR-pole}): they are at  $\alpha=\alpha_P$ (the pole of the beta function)
and at $\vert\alpha\vert=\infty$. Eq.(\ref{eq:RG-4loop-Pade}) then shows that  in the $k^2$ plane the  singularities
are all reached along the rays 

\begin{equation}k^2=\vert k^2\vert\exp\left(\pm {i\pi\over\gamma_{IR}}\right)\label{eq:ray-4loop}\end{equation}
where the phase arises from the imaginary part picked up by the first log on the right hand side of eq.(\ref{eq:RG-4loop-Pade}) when   
$\alpha=\alpha_P$ or $\vert\alpha\vert=\infty$. One deduces that 
the phases are larger then $\pi$, and therefore    the Landau singularities  are beyond the first sheet and the coupling
is causal
for $0<\gamma_{IR}<1$. In all cases we find the coupling is causal when the IR fixed point critical exponent is smaller then
one. A general explanation
of this fact is given in the next section.

\section{An equation to determine the bottom $N_f^*$ of the conformal window in QCD}
Let us assume from now on that the second scenario described in section 4 applies, i.e. that there is an IR fixed point
in the perturbative coupling even in some
range 
$N_f^c<N_f<N_f^*$ below $N_f^*$. To get a condition for causality breaking, one needs to know something on the location of Landau singularities.
It is clearly impossible to discuss all possible singularities  without
the knowledge of the full beta function.
I shall make the simplest assumption, namely that the Landau singularities originate only
from  the $\alpha>\alpha_{IR}$ region (some justification is provided below and in section 7), and argue that the condition

\begin{equation}0<\gamma<1\label{eq:gamma-causal}\end{equation}
is then  both necessary and sufficient for causality in QCD, where $\gamma$ is the critical exponent eq.(\ref{eq:exponent}). Consequently, the
lower boundary $N_f^*$  of the conformal window is obtained from the equation

\begin{equation}\gamma(N_f=N_f^*)=1\label{eq:gamma-bottom}\end{equation}
As is well known, the critical exponent
is a {\em universal} quantity, independent of the definition of the coupling, and  eq.(\ref{eq:gamma-bottom})
is a renormalization scheme invariant condition, as it should. 

 Assuming therefore  there is an 
$\alpha>\alpha_{IR}$  Landau singularity in the domain of attraction of $\alpha_{IR}$
(for instance a pole in the beta
function at $\alpha_P>\alpha_{IR}$ as in eq.(\ref{eq:UV-IR-pole})),
one first shows  \cite{Gar-Gru-conformal} that eq.(\ref{eq:gamma-causal})
is a 
necessary 
 condition for causality. 
I  give an improved version of the argument of \cite{Gar-Gru-conformal}. Solving the RG equation
$d\alpha/d\log k^2= \beta(\alpha)$
around $\alpha=\alpha_{IR}$, one gets
\begin{equation}\alpha(k^2)=\alpha_{IR}-\left({k^2\over\Lambda^2}\right)^{\gamma}+...
\label{eq:alpha-low}\end{equation}
There are thus rays
 \begin{equation}k^2=\vert k^2\vert\exp\left(\pm {i\pi\over\gamma}\right)\label{eq:ray}\end{equation} 
in the complex
 $k^2$ plane,
 which in the
infrared limit
$\vert k^2\vert\rightarrow 0$ are mapped by eq.(\ref{eq:alpha-low}) to positive real values of the coupling
{\em larger} then $\alpha_{IR}$. 
Assuming an expansion
\begin{equation}\beta(\alpha)=\gamma(\alpha-\alpha_{IR})+\gamma_1(\alpha-\alpha_{IR})^2+...
\label{eq:beta-low}\end{equation}
the corrections to eq.(\ref{eq:alpha-low}) are given by a series
\begin{equation}\log (k^2/\Lambda^2)={1\over \gamma} \log(\alpha_{IR}-\alpha)+{\gamma_1\over\gamma^2}
(\alpha_{IR}-\alpha)+...\label{eq:alpha-low-low}\end{equation}
with {\em real} coefficients,  showing that the only contribution to the phase for $\alpha>\alpha_{IR}$  comes from
the logarithm on the right hand side of eq.(\ref{eq:alpha-low-low}). The trajectories in the $k^2$ plane which
map  to the $\alpha>\alpha_{IR}$ region are thus straight lines to all
orders of perturbation theory around $\alpha_{IR}$. This fact suggests that
 even away from the infrared limit, 
these trajectories  are given by the rays eq.(\ref{eq:ray}). As $\vert k^2\vert$ is increased along these rays,
 the coupling will flow
  to the assumed 
Landau singularities, reached at some finite value of $\vert k^2\vert$. If $\gamma>1$ the 
 rays, hence also the  singularities,  are located on
the first sheet of the $k^2$ plane, showing that eq.(\ref{eq:gamma-causal}) 
is a necessary condition for causality. This condition  is also clearly
sufficient for causality, 
since I assume that
{\em no other sources} of (first sheet) Landau singularities
are present, but the one arising from the $\alpha>\alpha_{IR}$ region. 
A partial justification of the latter assumption shall  be provided in section 7.

That eq.(\ref{eq:gamma-causal}) is a necessary condition for causality is proved in Appendix A
 under the alternative assumption that
there is an $\alpha>\alpha_{IR}$ UV fixed point . This condition thus appears to be of quite general
validity.
It is interesting to note in this respect that a condition analogous to eq.(\ref{eq:gamma-causal})  has been derived \cite{Martin-Wells} 
from completely different
considerations as a consistency condition for non-asymptotically free supersymmetric gauge theories to have
 a non-trivial physical (positive) UV fixed point
(with $\gamma$ now being (minus) the critical exponent at the UV fixed point).

The very  existence of an $\alpha>\alpha_{IR}$ Landau singularity appears quite natural from another view point.
Indeed,
the alternative option  that there is an
$\alpha>\alpha_{IR}$ UV fixed point looks rather exotic, once one notices that  this 
fixed point may  still be present  in the non asymptotically free region $N_f>N_f^0$
(where the IR Banks-Zaks fixed point $\alpha_{IR}$
is no more physical, since it has moved to the $\alpha<0$ domain after vanishing at $N_f=N_f^0$), and would give a surprizing
example of a non-trivial, yet non asymptotically free field theory! On the other
hand, assuming a Landau singularity for $\alpha>\alpha_{IR}$ fits the standard expectation (``triviality'')
 that a (space-like)  Landau singularity is
present at positive coupling and is relevant to the non asymptotically free region above $N_f^0$. One thus gets the nice and economic picture
that essentially the {\em same} Landau singularity at $\alpha>0$ fits a dual purpose: below the conformal window
($N_f<N_f^*$) it provides the necessary  causality violation
and signals the emergence of a new non-perturbative phase, while
 above the conformal window ($N_f>N_f^0$) 
it is responsible for the
``triviality'' of the corresponding non-asymptotically free theory.

\section{Computing $N_f^*$ through the Banks-Zaks expansion}

 One can  try to use
 the Banks-Zaks
expansion \cite{BZ,White,G-bz} to compute $\gamma$ and determine $N_f^*$.
This  is an expansion of the fixed point in powers of the distance 
$N_f^0-N_f=16.5-N_f$
from the top of the conformal window, which is proportional to $\beta_0$. The solution of the equation

\begin{equation}\beta(\alpha)=-\beta_0 \alpha^2-\beta_1 \alpha^3-\beta_2 \alpha^4+...=0\label{eq:fixed-point}
\end{equation}
in the limit  $\beta_0\rightarrow 0$, with $\beta_i$ ($i\geq 1$) finite is obtained as a power series

\begin{equation}\alpha_{IR}=\epsilon+\left(\beta_{1,1}-{\beta_{2,0}\over\beta_{1,0}}\right)\epsilon^2+...\label{eq:BZ}
\end{equation}
where \cite{G-bz} the expansion parameter $\epsilon\equiv {8\over 321}(16.5-N_f)=
{\beta_0\over -\beta_{1,0}}={16\over 107}\beta_0$,  
$\beta_1\equiv\beta_{1,0}+\beta_{1,1}\ \beta_0$ and $\beta_{i,j}$ are $N_f$-independent (with $\beta_{i,0}$   
the $N_f=16.5$ values
of $\beta_i$).
The associated Banks-Zaks expansion for the critical exponent eq.(\ref{eq:exponent}) is presently known \cite{Stev,Gar-Kar} up 
to next-to-next-to leading order :

\begin{equation}\gamma={107\over 16}\epsilon^2(1+4.75 \epsilon-8.89 \epsilon^2+...)
\label{eq:exponent-BZ}\end{equation}
Using the truncated expansion eq.(\ref{eq:exponent-BZ}), one finds that $\gamma<1$ for $N_f\geq 5$,
with $\gamma=1$ reached for  $N_f=N_f^*\simeq 4$. To assess whether it is reasonable to use 
perturbation theory down to $N_f=N_f^*$, let us look at the magnitude of the successive terms within
the parenthesis in eq.(\ref{eq:exponent-BZ}). They are given by:
$1, 1.44, -0.82$. Although the next to leading term gives a very large correction, and the series
seem at best poorly converging at $N_f=N_f^*$, one can observe that the 
next-to next to leading term still gives a moderate correction to the sum of the first two terms, which
might be considered together \cite{Gar-Gru-conformal} as building the ``leading'' contribution,
 since they are
 both derived from information
contained \cite{G-bz,Gar-Gru-conformal} in the minimal 2-loop beta function necessary to get a non-trivial fixed point.
 Indeed, keeping only the first two terms
in eq.(\ref{eq:exponent-BZ}), one finds that $\gamma=1$ is reached for  $N_f=N_f^*\simeq 6$. On the
other hand, using a $[1,1]$ Pad\'e approximant as a model\footnote{The alternative $[0,2]$ Pad\'e
$\gamma={107\over 16}\ \epsilon^2/(1-4.75 \epsilon+31.45 \epsilon^2)$ 
 yields a result ($\gamma<0.26$ for all  real $N_f$)
 inconsistent with the present framework. It also predicts a not very plausible ${\cal O}(\epsilon^5)$
coefficient of $\simeq -192$.}
 for extrapolation of the perturbative
series, one gets

\begin{equation}\gamma={107\over 16}\ \epsilon^2\ {1+6.62 \epsilon\over 1+1.87 \epsilon}\label{eq:exponent-Pade}
\end{equation} 
which yields  $\gamma=1$ for $N_f=N_f^*\simeq 5$. The figure below shows  $\gamma$ as a function
of $N_f$:

\begin{figure}[H]
\begin{center}
\mbox{\kern-0.5cm
\epsfig{file=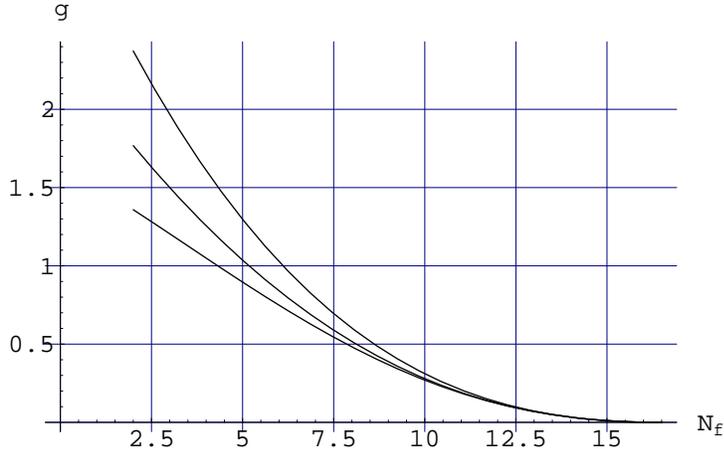,width=10.0truecm,angle=0}}
\end{center}
\caption{The critical exponent as a function of $N_f$:
 top: ${\cal O}(\epsilon^3)$ order; middle: Pad\'e; bottom: ${\cal O}(\epsilon^4)$ order.}
\label{pert}
\end{figure}

Note that in the obtained range of $N_f$ values
($4<N_f^*<6$),
$\beta_1$ is still positive ($\beta_1$ changes sign for $N_f\simeq 8$)
and of the same sign as $\beta_0$, so that the fixed point must arise
from the contributions of higher then 2 loop beta function corrections,
 although I am assuming the Banks-Zaks 
expansion is still converging there. This is consistent with the fact (see section 7) that many QCD effective charges have
 {\em negative} 3-loop beta function coefficients in the above range.

\section{Evidence for a negative UV fixed point in QCD}

Can one substantiate the crucial assumption which underlines the previous derivation that {\em no other sources} of 
Landau singularities are present, but the one arising from the $\alpha>\alpha_{IR}$ region?
 Even barring
{\em complex}
(in $\alpha$ space) Landau
 singularities (such 
as complex
poles in the beta function as in eq.(\ref{eq:complex-Landau}) below), a potential problem can still  arise from an 
eventual
  Landau 
singularity at $\alpha<0$,
in the domain of attraction of the {\em trivial} IR fixed point $\alpha=0^-$. In this section I make the important
additional  assumption
that perturbation theory is still asymptotic in the $\alpha<0$ region for the considered  beta function. This implies
that the corresponding coupling is itself free of renormalons ambiguities, despite their expected presence in generic
conformal window amplitudes (see section 8). Otherwise one would have to consider ambiguities suppressing contributions
(corresponding to power terms at short distances) which
most probably would induce essential singularities at zero coupling and blow up exponentially as $\alpha\rightarrow 0^-$. 
An attractive candidate would be the  ``skeleton coupling'' \cite{Brodsky} associated to an hypothetical
QCD ``skeleton expansion''.

Given this assumption, at weak coupling
the solution of the RG equation is controlled (even if $\alpha<0$) by the 2-loop beta function
\begin{equation}\log (k^2/\Lambda^2)={1\over \beta_0 \alpha}+{\beta_1\over \beta_0^2}\log \alpha+const+....
\label{eq:alpha-2loop}\end{equation}
where the $const$ is real. For $\alpha<0$ the right hand side of eq.(\ref{eq:alpha-2loop}) acquires a 
$\pm i\pi{\beta_1\over \beta_0^2}$ imaginary part, which implies   the rays
\begin{equation}k^2=\vert k^2\vert\exp\left(\pm i\pi{\beta_1\over \beta_0^2}\right)\label{eq:alpha<0}\end{equation}
map to the $\alpha<0$ region.
Along the rays eq.(\ref{eq:alpha<0}), we are effectively in a QED like situation:
increasing $\vert k^2\vert$, the coupling is either attracted to a non-trivial UV fixed point, or reaches a
Landau singularity at some finite $\vert k^2\vert$. In the latter case,
one must  require that
the condition:
\begin{equation}\vert{\beta_0^2\over \beta_1}\vert<1\label{eq:neg-landau-causal}
\end{equation}
 is satisfied in the whole $N_f$ range where eq.(\ref{eq:gamma-causal}) is valid, which 
will  confine  the rays, hence the singularities to the second (or higher) sheet.

It turns out that in QCD condition eq.(\ref{eq:neg-landau-causal}) can be satisfied
{\em only} if $\beta_1<0$,
and   coincides with the 2-loop causality condition 
eq.(\ref{eq:gamma-2loop}), 
  which requires $N_f>9.7$. Eq.(\ref{eq:neg-landau-causal}) is however
 a necessary condition for causality for {\em any} beta function which admits a Landau singularity at negative $\alpha$
 (in the domain
of attraction of the $0^-$ trivial IR fixed point),
 and applies also if $\beta_1$
is positive in a general theory!
Therefore, to preserve causality within
 the conformal window as determined
 by eq.(\ref{eq:gamma-causal}),  $\gamma$ should reach $1$ in the region $N_f>9.7$,
 which is clearly excluded
 (see Fig. 1). In order that eq.(\ref{eq:gamma-causal}) be also a sufficient condition for causality
one must thus check   that a {\em non-trivial}
(finite or infinite) UV 
fixed point $\alpha_{UV}$ is present 
at {\em negative} $\alpha$.
A minimal example satisfying this requirement is the  3-loop beta function 
eq.(\ref{eq:3-loop})
with $\beta_0>0$ and $\beta_2<0$ ($\beta_1$ can have any sign). Another example
is provided by eq.(\ref{eq:UV-IR-pole}).

It is  worth mentioning  eq.(\ref{eq:neg-landau-causal}) is
{\em always}
violated \cite{Gar-Gru-conformal} in the lower part of the conformal window in  SQCD as determined by duality \cite{Seiberg},
 and the 
previous argument
 thus implies the existence
 of a negative UV fixed point in this theory. In fact the ``exact'' NSVZ \cite{NSVZ} beta function for $N_f=0$

\begin{equation}\beta(\alpha)=-{\beta_0 \alpha^2\over 1- {\beta_1\over \beta_0}\alpha}\label{eq:NSVZ}\end{equation}
 does exhibit  an (infinite) UV fixed point as $\alpha\rightarrow -\infty$, which might
be the parent of a similar one present within Seiberg conformal window \cite{Seiberg}.

It is a priori possible in QCD to have an $\alpha<0$  Landau singularity 
rather then an $\alpha<0$ UV fixed point.  A simple example (apart from
 the two-loop beta function eq.(\ref{eq:2-loop}) with $\beta_0>0$ and $\beta_1<0$) is the three loop
beta function eq.(\ref{eq:3-loop}) with $\beta_0, \beta_2>0$ and $\beta_1<0$, with a {\em positive} UV fixed point
 $\alpha_{UV}>\alpha_{IR}$ (one has to assume that the IR fixed point does not disappear {\em before} the
$\alpha<0$  Landau singularity appears on the first sheet, i.e. the reverse situation to the one considered 
after eq.(\ref{eq:3-loop})).
In such a case
the bottom of the conformal window would be given as in the two-loop model by the condition $-\beta_0^2/\beta_1=1$, yielding
$N_f^*\simeq 9.7$. Indeed, at such large $N_f$, any eventual  Landau singularity from the $\alpha>\alpha_{IR}$ region has not
yet reached the first sheet (as already observed) since $\gamma<0.4$  as shown by Fig. 1. This possibility is however disfavored
 as now explained.

\underline{A modified Banks-Zaks expansion}: 
there is indeed some evidence that a {\em negative} UV fixed point is actually  present
 in QCD. At the three-loop level, QCD effective
charges associated to Euclidean correlators typically \cite{Gar-Kar,Gar-Gru-conformal}
have  $\beta_2<0$, and
 appear  \cite{Gar-Gru-conformal} to be causal and  admit a negative
UV fixed point,  even somewhat {\em below} the two-loop
causality boundary $N_f = 9.7$ (in a range $7.0<N_f<9.7$). This evidence could be washed out in yet higher orders (e.g. the Pad\'e improved
three-loop beta functions \cite{Gar-Gru-conformal}). However, additional\footnote{Yet another evidence is provided by the
 observation that $\gamma$
  (eq.(\ref{eq:exponent-BZ})
or(\ref{eq:exponent-Pade})) vanishes for $\epsilon\simeq -0.15$ (where the Banks-Zaks expansion is still convergent!),
 i.e. for $N_f\simeq 22.6$ and can be interpreted as the point where
$\alpha_{UV}$ coincides with $\alpha_{IR}$ (which is negative for $N_f>16.5$).} and more systematic
 evidence for the existence
of a {\em couple} of (positive-negative) IR-UV fixed points  is provided by the following modified Banks-Zaks argument.
Assume $\beta_1=0$, i.e. $N_f=8.05$.  Then a real
 fixed point can
 still exist at the three-loop level
if $\beta_2<0$, and actually one gets a {\em pair} of opposite signs zeroes at ${\tilde \alpha}= \pm (-\beta_0/\beta_2)^{1/2}$,
 an IR
and an UV fixed point. If $\beta_0$ is small enough, they are weakly coupled, and calculable through a modified
Banks-Zaks expansion around $N_f=16.5$,  applied to the auxiliary function
 ${\tilde \beta}(\alpha)\equiv  \beta(\alpha)+\beta_1\alpha^3$ with the two-loop term removed. One gets

\begin{equation}{\tilde \alpha}_{IR,UV}=\pm {\tilde \epsilon}\left(1 \mp {1\over 2}\ {\beta_{3,0}\over \beta_{2,0}}\ {\tilde \epsilon}+...\right)
\label{eq:BZ2}\end{equation} 
where the expansion parameter ${\tilde \epsilon}\equiv (-\beta_0/\beta_{2,0})^{1/2}$, and $\beta_{i,0}$,  
the $N_f=16.5$ values
of $\beta_i$ ($i=2,3$), are scheme dependent. Given a reference scheme  whose beta function is known up to 4-loop
(say, the 
$\overline {MS}$ scheme),
$\beta_{2,0}$ can be obtained \cite{G-BLM} from the $N_f=16.5$ value
of the  next-to-leading order coefficient in the expansion of the considered coupling in this scheme. Furthermore
 $\beta_{3,0}$ follows from
the knowledge of 
the next-to-leading  ($g_1=\beta_{1,1}=4.75$) and next-to-next-to-leading order ($g_2=-8.89$) coefficients
 in eq.(\ref{eq:exponent-BZ}) and
 the relation \cite{G-bz,Gar-Kar} (which yields $g_2$ to start with  when used \cite{Stev} in the reference scheme, since 
the $g_i$'s are scheme invariant \cite{G-bz}) 
\begin{equation}{16\over 107}\ \beta_{3,0}+\left({16\over 107}\ \beta_{2,0}\right)^2=g_1^2-g_2 \sim 31.4\label{eq:BZ3}\end{equation}
 For  effective charges associated to Euclidean correlators,
 the correction in eq.(\ref{eq:BZ2}) ranges 
from $0.1$ to $0.7$ at $N_f=8.05$ (where ${\tilde \beta}(\alpha)$ coincides with  $ \beta(\alpha)$),
 and gives some
evidence for a couple of UV and IR fixed points around this value of $N_f$ (which is within the alleged
conformal window, but below the 2-loop causality region). However, as far as the negative fixed point is concerned,
 this evidence is corroborated only if one assumes 
asymptoticity of perturbation theory at negative coupling for the considered beta functions. As pointed out above, this can hardly be expected
for the beta functions of physical effective charges such as those associated to Euclidean correlators. The modified
Banks-Zaks expansion should rather be applied to the aforementioned  perturbative ``skeleton coupling'',
 which may be free of renormalons
and of the associated power terms. Assuming the correct ``skeleton coupling'' is the one identified \cite{Watson,Papavassiliou}
 through the 
``pinch technique'', one gets \cite{Brodsky} $\beta_{2,0}=-17.5$
and eq.(\ref{eq:BZ3}) yields $\beta_{3,0}=164.7$. Unfortunately, for these values\footnote{The skeleton coupling has the
{\em smaller} value of $\vert\beta_{2,0}\vert$, resulting (eq.(\ref{eq:BZ3})) into the {\em larger} value of $\beta_{3,0}$ and of the correction term
 in eq.(\ref{eq:BZ2})!}
the correction factor
 in eq.(\ref{eq:BZ2}) is $1.34$
at $N_f=8.05$ and the expansion diverges! Nevertheless, these large corrections may
  indicate the necessity of resummation of the modified Banks-Zaks expansion
rather then the absence of the fixed points. Some support for this interpretation  is indicated by the satisfactory
 convergence 
  of the {\em standard} (eq.(\ref{eq:BZ})) Banks-Zaks expansion  for the IR fixed point of the skeleton coupling

\begin{equation}\alpha_{IR}=\epsilon+2.14\epsilon^2+...\label{eq:BZ-skel}
\end{equation}
which yields
 a next-to-leading order correction factor \cite{G-bz} of about $0.45$ at $N_f=8.05$ (for the Euclidean
 effective charges, the convergence is even better \cite{Stev,Gar-Kar,Gar-Gru-conformal,Brodsky}). It is also possible to 
investigate the existence of a negative UV fixed point at 
higher values of $N_f$ (where $\beta_1$ does not vanish), using a generalization of the modified Banks-Zaks expansion 
(see Appendix B). One finds  slightly
improved next-to-leading order corrections (of about $80\%$) at $N_f=11$. This makes the existence of a couple of
 (IR,UV) fixed points at least
 plausible
 around this value
of $N_f$. However, below $N_f=9.7$, i.e. below the 2-loop causality region (where
 the negative UV fixed point is really needed), the corrections to the modified Banks-Zaks fixed point series are still 
over $100\%$.

Additional support is given by the calculation 
of the auxiliary critical exponent
${\tilde \gamma}\equiv
d{\tilde\beta}(\alpha)/ d\alpha\vert_{\alpha={\tilde\alpha_{IR}}}$, which gives

\begin{equation}{\tilde \gamma}= 2\ \beta_0\ {\tilde \epsilon}(1+{\cal O}({\tilde \epsilon}^2))\label{eq:BZ2-exponent}\end{equation}
(no ${\cal O}({\tilde \epsilon})$ correction!). For  effective charges associated to Euclidean correlators one gets
 $0.6<{\tilde \gamma}<0.7$ at $N_f=8.05$ (where ${\tilde \gamma}$ coincides with  $\gamma$), in reasonable agreement with the standard
 Banks-Zaks result (Fig. 1) $0.5<\gamma<0.6$. The agreement is again less satisfactory for the skeleton coupling, which
yields ${\tilde \gamma}\sim 0.8$, but still qualitatively acceptable indicating  the need for resummation.

It should be stressed that the non-trivial negative UV fixed point is actually {\em not relevant} to the proper analytic continuation of
 the 
coupling at complex $k^2$, which must be consistent \cite{Gar-Gru-Kar} with (UV) asymptotic freedom. The latter condition
ensures that  eventual first sheet Landau singularities are localized within a bounded infrared region, as expected on physical
grounds.
 This means that 
in presence
of this fixed point, the correct analytic continuation must involve {\em complex} rather then negative
values of $\alpha$ along the rays eq.(\ref{eq:alpha<0}), and one should  approach
 the non-trivial
(rather then the trivial) IR fixed point as $\vert k^2\vert\rightarrow 0$, 
and the trivial (rather then the non-trivial) UV fixed point as $\vert k^2\vert\rightarrow \infty$. This is possible since the
solution of eq.(\ref{eq:alpha-2loop}) is not unique for a given (complex) $k^2$.
 For a similar
 reason, any eventual 
Landau singularity arising from the region $\alpha<\alpha_{UV}$, in the domain of attraction
of the non-trivial UV fixed point, is   not relevant to the correct analytic continuation. However, I have to
 assume\footnote{A counterexample is given by the 4-loop beta function
 $\beta(\alpha)=-\beta_0 \alpha^2-\beta_1 \alpha^3-\beta_2 \alpha^4-\beta_3 \alpha^5$ with one IR and two UV fixed points
such that $\alpha_{UV}^{'}<0<\alpha_{IR}<\alpha_{UV}$. In this example, the Landau singularities are at
 $\vert\alpha\vert=\infty$ ($\alpha$ complex)
and cannot be reached neither from the $\alpha_{IR}<\alpha<\alpha_{UV}$ nor from the $\alpha_{UV}^{'}<\alpha<0$ regions.}
that the considered beta function is such that all  $\alpha>\alpha_{IR}$ singularities (and more generally all
 $\vert\alpha\vert=\infty$ singularities) 
are either irrelevant, or  have the same phase $\pi/\gamma$ in $k^2$ plane, as in 
the example eq.(\ref{eq:UV-IR-pole}).
 
On the other hand,
 it was implicitly
assumed above  that
 any $\alpha<0$ Landau
singularity in the domain of attraction of  the trivial $0^-$ IR fixed point
  {\em is relevant}, i.e. that the coupling will flow to the 
   trivial
$0^+$ UV fixed point once the $\alpha<0$ singularity is passed, even if there is another (positive) UV fixed point, as in the 
three loop example below eq.(\ref{eq:NSVZ}). Similarly, in presence of the negative UV fixed point, one has
to {\em assume}  that the coupling will still flow to the 
trivial UV fixed point (rather then  the negative one), once the  $\alpha>\alpha_{IR}$  Landau
 singularity is passed. This must be the case
for consistency of the present approach, where causality violation in a coupling which satisfies UV
asymptotic freedom is attributed either to the $\alpha<0$ or
to the $\alpha>\alpha_{IR}$ Landau singularities.

\section{``Perturbative'' and ``non-perturbative'' components of condensates: the two loop APT coupling as a toy
 ``non-perturbative'' model}

To illustrate the  discussion in section 3, consider as a model for the full QCD amplitude $D(Q^2)$ 
 the standard ``renormalon integral''

\begin{equation} D(Q^2)
=  \int_0^{Q^2}{dk^2\over k^2}\ \bar{\alpha}(k^2)\
n\ \left({k^2\over Q^2}\right)^n\label{eq:renormalon-integral}\end{equation}
where the ``non-perturbative'' coupling $\bar{\alpha}(k^2)$ (tentatively identified as a non-perturbative extension of the ``skeleton coupling''
\cite{Brodsky}, as suggested \cite{Grunberg-power,Gar-Gru-thrust} by the IR finite coupling approach \cite{Dok} to power corrections) is IR finite and causal. I further  assume the perturbative part
 $\alpha(k^2)$ of the coupling  
is  by itself IR finite even  below the conformal window, but still differs there (where
 it is not causal)  from the full
  non-perturbative coupling $\bar{\alpha}(k^2)=\alpha(k^2)+\delta\alpha(k^2)$.
Thus  in this model eq.(\ref{eq:PT-NP}) holds, with

\begin{equation}D_{\overline{PT}}(Q^2)\equiv \int_0^{Q^2}{dk^2\over k^2}\ \alpha(k^2)\
n\ \left({k^2\over Q^2}\right)^n\label{eq:PT-bar}\end{equation}
and

\begin{equation}D_{\overline{NP}}(Q^2)\equiv \int_0^{Q^2}{dk^2\over k^2}\ \delta\alpha(k^2)\
n\ \left({k^2\over Q^2}\right)^n\label{eq:NP-bar}\end{equation}
One can show \cite{Grunberg-FP,D-U} that $D_{\overline{PT}}(Q^2)$
differs from the corresponding Borel sum $D_{PT}(Q^2)$

\begin{equation}D_{PT}(Q^2)\equiv \int_0^\infty dz \exp\left(-z/\alpha(Q^2)\right)
 \tilde D(z)\label{eq:Borel-sum}
\end{equation}
by a power correction
\begin{equation} D_{\overline{PT}}(Q^2)=D_{PT}(Q^2)+[\tilde C_{PT} (-1)^{\delta}+C_{PT}]
\left({\Lambda^2\over Q^2}\right)^n\label{eq:PT-power-corr}\end{equation}
where $\delta=n{\beta_1\over\beta_0^2}$ and
 $C_{PT}$ and $\tilde C_{PT}$ are real constants. $\tilde C_{PT}$ is the
normalization  of a complex component,
 which cancels
the ambiguity of $D_{PT}(Q^2)$ due to IR renormalons. Both $C_{PT}$ and $\tilde C_{PT}$ can in principle  be
 determined \cite{Grunberg-FP,D-U}
from information contained in the (all-orders) perturbative series which build up $D_{PT}(Q^2)$ (see also below and Appendix C).
They represent the ``perturbative'' part of the condensate, the only one surviving within the conformal window.
On the other hand,
if $\delta\alpha(k^2)$ decreases faster then
${\cal O}(1/k^{2n})$ at large $k^2$ (so that the leading power correction is of IR origin, i.e. the
integral eq.(\ref{eq:NP-bar})  is not dominated by its UV tail), one gets for
$Q^2\gg\Lambda^2$

\begin{equation} D_{\overline{NP}}(Q^2)\simeq C_{NP}
\left({\Lambda^2\over Q^2}\right)^n+...\label{eq:NP-power-corr}\end{equation}
with
\begin{equation}C_{NP}=\int_0^\infty{dk^2\over k^2}\ \delta\alpha(k^2)\
n\ \left({k^2\over \Lambda^2}\right)^n\label{eq:C-NP}\end{equation}
which represents the ``genuine'' non-perturbative part of the condensate, and should
vanish within the conformal window. Thus for $Q^2\gg\Lambda^2$

\begin{equation} D(Q^2)\simeq D_{PT}(Q^2)+C
\left({\Lambda^2\over Q^2}\right)^n+...\label{eq:power-corr}\end{equation} 
with 
\begin{equation}C=\tilde C_{PT} (-1)^{\delta}+C_{PT}+C_{NP}\label{eq:C-PT-NP}\end{equation}

Let us  now take the  ``APT coupling'' $\alpha_{APT}(k^2)$
 as a toy model for the full
 non-perturbative
coupling $\bar{\alpha}(k^2)$ below the conformal window. It is  defined \cite{Shirkov} by the dispersion relation

\begin{equation}\alpha_{APT}(k^2)
\equiv  -\int_0^\infty{d\mu^2\over
\mu^2+k^2}\
\rho(\mu^2)\label{eq:APT}\end{equation}
where  the ``spectral density''

\begin{equation}\rho(\mu^2) = \frac{1} {2\pi i}
 Disc\{\alpha(-\mu^2)\}
\equiv -\frac{1}
{2\pi i}\{\alpha\left[-(\mu^2+i\epsilon)\right]-\alpha
\left[-(\mu^2-i\epsilon)\right]\} \label{eq:disc}\end{equation}
 is proportional to
the time-like discontinuity  of the perturbative coupling $\alpha(k^2)$. Eq.(\ref{eq:APT})
 implies in particular the absence of (first sheet) Landau singularities, and is therefore a 
``brute force'' causal deformation of the perturbative coupling $\alpha(k^2)$ (within the conformal window, where
 $\alpha(k^2)$ is causal,
it coincides with $\alpha_{APT}(k^2)$ ). At large $k^2$, the discrepancy
$\delta\alpha_{APT}(k^2)$  between
 $\alpha_{APT}(k^2)$ and $\alpha(k^2)$ is an ${\cal O}(1/k^2)$ power correction, and I
therefore assume $0<n<1$ to guarantee the leading power correction is of IR origin. 
One then finds at large $Q^2$ that

\begin{equation} D(Q^2)\equiv D_{APT}(Q^2)\simeq D_{PT}(Q^2)+C_{APT}
\left({\Lambda^2\over Q^2}\right)^n +...\label{eq:power-corr-APT}\end{equation}
where, assuming from now on that $\alpha(k^2)$ is the {\em two-loop} coupling eq.(\ref{eq:2-loop}),
 the Borel sum $D_{PT}(Q^2)$ is given by \cite{Grunberg-FP,D-U}

\begin{equation} D_{PT}(Q^2)=\int_0^\infty dz \exp\left(-z/\alpha(Q^2)\right)
 {\exp\left(-{\beta_1\over\beta_0}z\right)\over\left(1-{z\over z_n}\right)^{1+\delta}}\label{eq:Borel}
\end{equation}
while for the power term one finds

\begin{equation}C_{APT}=(-1)^n{\pi\over\sin\pi
n}{z_n^{1+\delta}\over\Gamma(1+\delta)}\label{eq:C-APT}
\end{equation}
where $z_n=n/\beta_0$
and $\Lambda$ is defined from the solution of the two-loop renormalization group equation

\begin{equation}\beta_0\log{Q^2\over\Lambda^2}={1\over\alpha}-{\beta_1\over\beta_0}
\log\left({1\over\alpha}+{\beta_1\over\beta_0}\right)+{\beta_1\over\beta_0}\label{eq:RG}\end{equation}
The result eq.(\ref{eq:C-APT}) is obtained immediately from a similar one in \cite{Gar-Gru-thrust},
taking into account the different definition of $\Lambda$ used here (which is not the Landau singularity). This result
actually assumes a space-like Landau singularity, i.e. that $\beta_1>0$, but eq.(\ref{eq:C-APT}) makes also sense 
 for $\beta_1<0$
where there is an IR fixed point, which suggests it might apply in this case as well, provided there are {\em
complex} Landau singularities and causality is violated\footnote{When the coupling is causal, i.e. for
$-{\beta_1\over\beta_0^2}>1$, $\alpha$ and $\alpha_{APT}$ coincide as mentioned above, and $C_{APT}$ is given 
instead
by $\tilde C_{PT}$  (eq.(\ref{eq:C-tilde-PT})).}, i.e. for 

\begin{equation}0<-{\beta_1\over\beta_0^2}<1\label{eq:2-loop-noncausal}\end{equation}
Moreover  when $\beta_1<0$
  $D_{\overline{PT}}(Q^2)$ (eq.(\ref{eq:PT-bar}))
is given by \cite{Grunberg-FP,D-U}

\begin{eqnarray}D_{\overline{PT}}(Q^2)&=& \int_0^{z_n} dz \exp\left(-z/\alpha(Q^2)\right)
 {\exp\left(-{\beta_1\over\beta_0}z\right)\over\left(1-{z\over z_n}\right)^{1+\delta}}\nonumber\\
&=& \int_0^\infty dz \exp\left(-z/\alpha(Q^2)\right)
 {\exp\left(-{\beta_1\over\beta_0}z\right)\over\left(1-{z\over z_n}\right)^{1+\delta}}
-\int_{z_n}^\infty dz \exp\left(-z/\alpha(Q^2)\right)
 {\exp\left(-{\beta_1\over\beta_0}z\right)\over\left(1-{z\over z_n}\right)^{1+\delta}}\nonumber\\
 &\equiv& D_{PT}(Q^2)+\tilde C_{PT} (-1)^{\delta}\left({\Lambda^2\over
Q^2}\right)^n\label{eq:power-corr-PT}\end{eqnarray}
which shows that in this 2-loop example  the ``perturbative part'' of the power correction coincides with the ``renormalon
part'', i.e. we have  in eq.(\ref{eq:PT-power-corr})

\begin{equation}C_{PT}=0\label{eq:C-PT}\end{equation}
while \cite{Grunberg-FP,D-U}

\begin{equation}\tilde C_{PT}=-{\pi\over\sin\pi\delta} {z_n^{1+\delta}\over
\Gamma(1+\delta)}\label{eq:C-tilde-PT}\end{equation}
One can thus write eq.(\ref{eq:C-APT}) in the form of eq.(\ref{eq:C-PT-NP}), i.e. we have 
$C_{APT}=\tilde C_{PT} (-1)^{\delta}+C_{PT}+C_{NP}$ with $\tilde C_{PT}$ and $C_{PT}$ as above,
and one also finds

\begin{equation}C_{NP}=\pi{z_n^{1+\delta}\over\Gamma(1+\delta)}\left({\cos\pi n\over\sin\pi n}+
{\cos\pi \delta\over\sin\pi \delta}\right)\label{eq:C-NP-APT}\end{equation}
It is interesting to note that $C_{NP}$  vanishes when $\delta=-n$ (remember $0<n<1$),
 i.e. for
${\beta_1\over \beta_0^2}=-1$, which is the bottom of the conformal window in this model. This fact can be
understood since there one might
 expect\footnote{This expectation is actually not realized beyond 2-loop, where one finds \cite{Gar-Gru-Kar}
 there is no continuity between the IR values of $\alpha_{APT}$ above and below the conformal window.} (in the 2-loop case) $\alpha_{APT}=\alpha$, which is an indication in favor of
the correctness of eq.(\ref{eq:C-APT}) when $\beta_1$ satisfies eq.(\ref{eq:2-loop-noncausal}).
 $C_{NP}$ does not however vanish
identically within the conformal window, which means that
$C_{NP}$ can be analytically continued from below  to within the conformal window, but 
 does not
give the correct power correction there (see  footnote 8).

\section{A rationale for the infrared finite coupling approach to power corrections}

The values of $N_f^*$ obtained in section 6 are substantially {\em lower} then the one following from
the naive 2-loop causality condition ($N_f^*=9.7$) or the similar one  from an earlier
attempt based on ``superconvergence''  \cite{Gar-Gru-conformal}, and rather close to the  number of
 light
QCD flavors $N_f=3$ usually relevant for QCD phenomenology. As I now argue, this fact may give 
the underlying justification to the IR finite coupling approach \cite{Dok} to power corrections.
Indeed, the very notion that there is a non-perturbative modification $\delta\alpha$  of the
coupling (section 8) may be only a convenient phenomenological parametrization of power corrections, without
fundamental meaning. On the other hand, the existence of an IR finite coupling appears very 
natural within the conformal window, but there it is of entirely perturbative origin, with no need
for a  $\delta\alpha$ term. The following picture then suggests itself as an attractive alternative: in the decomposition
eq.(\ref{eq:PT-NP}), only the 
``perturbative'' conformal window amplitude $D_{\overline{PT}}$ may be related to the IR finite universal
 coupling (see e.g.
eq.(\ref{eq:PT-bar})), and there is no such thing as a non-perturbative 
modification of the coupling (a $\delta\alpha$ term). Still at large $Q^2$ one can write

\begin{equation} D(Q^2)\simeq D_{\overline{PT}}(Q^2)+C_{NP}
\left({\Lambda^2\over Q^2}\right)^n+...\label{eq:power-corr-NP}\end{equation}
with $C_{NP}$ now an independent parameter characterizing the ``non-perturbative'' vacuum,
and bearing no connection (such as  eq.(\ref{eq:C-NP})) to the universal coupling $\alpha$. The consecutive lost of predictive power
 following a priori from these more general assumptions is however compensated by the observation that
$C_{NP}$ vanishes by definition for $N_f\geq N_f^*$, and can be expected to be {\em still small} for
$N_f$ {\em not too far below} $N_f^*$, if the phase transition is ``second order'', i.e. if $C_{NP}$  is continuous as a function
of $N_f$, and vanishes as $N_f$ approaches $N_f^*$ from {\em below} (as in the 2-loop APT model of 
eq.(\ref{eq:C-NP-APT})). It is thus at least plausible,
 given the low range $4<N_f^*<6$, that even at $N_f=3$ one can still neglect $C_{NP}$, i.e. that the
main contribution to the condensate is given by its ``perturbative'' conformal window piece
$\tilde C_{PT} (-1)^{\delta}+C_{PT}$ (see eq.(\ref{eq:C-PT-NP})). An even more radical possibility
would be that condensates are either entirely ``perturbative'' or ``non-perturbative'', i.e. that
 $C_{NP}\equiv 0$ even {\em below} $N_f^*$ for those condensates which do not vanish within the
conformal window. In such a case, the IR finite coupling approach would be justified at all $N_f$'s,
independently of the value of $N_f^*$, as long as an IR fixed point is still present in the
 {\em perturbative} coupling.

This view raises the interesting
 possibility that the main contribution to the condensate may  actually be {\em calculable} by 
{\em perturbative} means. To see this, recall \cite{Dok} that power corrections can be
parametrized in terms of low energy moments of the IR finite coupling. For instance in 
eq.(\ref{eq:PT-bar}) the low energy part below an IR cut-off $\mu_I$ yields a power correction
$\lambda(\mu_I)/Q^{2n}$ with $\lambda(\mu_I)=\int_0^{\mu_I^2}{dk^2\over k^2}\ \alpha(k^2)\
n\ k^{2n}$. As a first rough  estimate (a more precise calculation shall be presented elsewhere
\cite {Gru-next}) one can replace  $\alpha(k^2)$ by $\alpha_{IR}$ in the
expression for $\lambda(\mu_I)$, and use the Banks-Zaks expansion eq.(\ref{eq:BZ-skel}) to evaluate
$\alpha_{IR}$ in the skeleton scheme (identified to the pinch technique coupling). In the range $4<N_f<6$ corresponding to the bottom of the conformal
window one finds\footnote{These values are compatible with 
the relation $\beta_0\ \alpha_{IR}=1$ at the bottom of the conformal window, which is
 the ``universal'' IR value of the non-perturbative APT
coupling \cite{Shirkov} (and follows also from the 2-loop model eq.(\ref{eq:gamma-2loop})).
 The significance of this observation is however
 unclear, given the large corrections involved.} $0.41<\alpha_{IR}<0.52$, while at $N_f=3$ 
one finds $\alpha_{IR}=0.58$, with next to leading order corrections of the order
60-70\% (I defined $\alpha\equiv \alpha_s/\pi$). These corrections, while substantial, do not completely rule out a perturbative
approach.
To get an hopefully more reliable estimate of 
$\alpha_{IR}$ at the bottom of the conformal window, one can indeed
 eliminate the
Banks-Zaks parameter $\epsilon$ in eq.(\ref{eq:exponent-BZ}) in favor of $\alpha_{IR}$ using
 eq.(\ref{eq:BZ-skel}) to
obtain

\begin{equation}\gamma={107\over 16}\alpha_{IR}^2(1+0.47 \alpha_{IR}+...)
\label{eq:exponent-BZ-improved}\end{equation}
which seems to converge better. In leading order, one  obtains $\gamma=1$ for 
$\alpha_{IR}=(16/107)^{1/2}\simeq 0.39$ (a scheme invariant prediction), whereas in next-to-leading order
$\gamma=1$ is reached for   $\alpha_{IR}\simeq 0.36$, with a small next-to-leading order  correction
of 0.17.

The possible relevance of 
conformal window physics and of the ``perturbative'' Banks-Zaks freezing of the coupling at low $N_f$
suggested here is 
 very specific, and applies only to the calculation of ``condensates'' (already present within the
conformal window) which appear in the short distance contributions to amplitudes. In particular,
it is not assumed that the whole ``non-perturbative'' component $D_{\overline{NP}}(Q^2)$ in 
eq.(\ref{eq:PT-NP})
is small for any $Q^2$: this is clearly not correct at $Q^2=0$ for the effective
charge associated to the Adler D-function, whose exact value there  is known  from spontaneous chiral symmetry breaking arguments \cite{Rafael}  and turns out to be negative, thus inconsistent\footnote{This discrepancy
can be understood as the effect of the decoupling of all quarks flavors at $Q^2=0$, due to their
non-vanishing dynamical mass from (spontaneous) chiral symmetry breaking below the conformal
window. Thus at $Q^2=0$ one is effectively in an $N_f=0$ theory, and never close to the bottom
of the conformal window.}  with the positive
result \cite{Stev} following from the Banks-Zaks expansion at $N_f=3$. Thus 
 the present suggestion differs from the related ideas in \cite{Stev}.

\section{Behavior of the conformal window amplitudes in the 
$N_f\rightarrow -\infty$ ``one-loop'' limit}
Although the arguments in section 5 require an IR fixed point in the perturbative coupling to be present
 only in some {\em finite} range
$N_f^c<N_f<N_f^*$ below $N_f^*$, it is attractive to consider the possibility that
 the IR fixed point may actually persists down to the $N_f\rightarrow -\infty$ limit (where the ``skeleton
coupling'' becomes \cite{Brodsky} one loop). In such a case, the natural question arises whether the full conformal window amplitudes, which
contain power terms, stay finite
 or blow up 
 in this limit (despite the perturbative series coefficients and the corresponding Borel sum remaining themselves finite).
 That the latter  can
 easily happen is demonstrated by the simplest 2-loop example eq.(\ref{eq:C-tilde-PT}), which shows that $\tilde C_{PT}$ is singular
 for $\delta\rightarrow 0$, i.e. in the  $\beta_1\rightarrow 0$ one-loop limit, while the Borel sum
 itself (eq.(\ref{eq:Borel})) is finite. The singular behavior of $\tilde C_{PT}$ 
is actually a general fact. Indeed, the contribution of a dimension $2 n$ operator in the OPE can always be parametrized 
\cite{Gru-condensate} (assuming no anomalous dimension)
 as in eq.(\ref{eq:power-corr}) and (\ref{eq:C-PT-NP}), where the $\tilde C_{PT}$ component is the
``renormalon contribution'' given by (minus) the singular part for $z\rightarrow z_n$ of the Borel transform

\begin{eqnarray} \tilde C_{PT} (-1)^\delta \left({\Lambda^2\over Q^2}\right)^n&=&\tilde C_{PT} \exp(-\delta) 
\exp\left(-z_n/\alpha(Q^2)\right) (-\alpha(Q^2))^{-\delta}\left[1+{\cal O}(\alpha(Q^2))\right]\nonumber\\
 &=&-\int_{z_n}^\infty dz \exp\left(-z/\alpha(Q^2)\right)
 {K\over\left(1-{z\over z_n}\right)^{1+\delta}}\left[1+{\cal O}\left(1-{z\over z_n}\right)\right]\label{eq:ren.part}
\end{eqnarray}
where  $\delta=n{\beta_1\over\beta_0^2}$  (as below eq.(\ref{eq:PT-power-corr})) and the factor $\exp(-\delta)$ is due to 
the definition of $\Lambda$ eq.(\ref{eq:RG}).
  $K$ is the renormalon residue,
 related to $\tilde C_{PT}$ by

\begin{equation} \tilde C_{PT}=-K \exp(\delta) {\pi\over \sin \pi\delta}{z_n^{1+\delta}\over\Gamma(1+\delta)}\label{eq:C-K}
\end{equation} 
 Note that eq.(\ref {eq:C-K}) reproduces
 eq.(\ref{eq:C-tilde-PT}) for $K=\exp(-\delta)$,
 which is indeed the correct renormalon residue in this case (see eq.(\ref{eq:Borel})). In the one-loop limit where
 $\beta_1, \delta \rightarrow 0$,
one finds the singular behavior (assuming $K$ remains finite as suggested by the behavior of  the perturbative series
 coefficients)

\begin{equation} \tilde C_{PT}\simeq - K {z_n\over\delta}=-K {\beta_0\over \beta_1}\label{eq:C-sing}
\end{equation}
It does not mean of course the full amplitude is singular, since there can be a cancellation 
between $\tilde C_{PT} (-1)^{\delta}$ and $C_{PT}+C_{NP}$ in eq.(\ref{eq:C-PT-NP}). The interesting question however is 
whether the perturbative ($\tilde C_{PT} (-1)^{\delta}+C_{PT}$) and non perturbative ($C_{NP}$) components of the condensate
 remain {\em separately} finite, i.e. the cancellation takes place between $\tilde C_{PT}$ and $C_{PT}$, or whether
 the cancellation (if any\footnote{I am actually not aware of any physical reason
why the full amplitude should behave as its perturbative series, i.e. remain finite in the
$N_f\rightarrow -\infty$ limit.})
takes place between the perturbative and the non perturbative pieces? If the perturbative condensate remains finite,
eq.(\ref{eq:PT-power-corr}) and (\ref{eq:Borel}) become in the one-loop limit

\begin{equation}D_{\overline{PT}}(Q^2)=D_{PT}(Q^2)+(const\pm i\pi  z_n)
\left({\Lambda^2\over Q^2}\right)^n \label{eq:one-loop-limit}
\end{equation}
with

\begin{equation} D_{PT}(Q^2)=\int_0^\infty dz \exp\left(-z/\alpha(Q^2)\right)
 {1\over1-{z\over z_n}}\label{eq:Borel-oneloop}
\end{equation}
where the imaginary part in the power term removes the one-loop renormalon ambiguity in eq.(\ref{eq:Borel-oneloop}).

 Cancellation between the perturbative and non-perturbative components of the condensate occurs in the two-loop 
 APT model of section 8, where $C_{PT}=0$ (and thus cannot cancel the divergent $\tilde C_{PT}$)
 while  $C_{APT}$ is finite
(see eq.(\ref{eq:C-APT})). On the other hand, in the three-loop case eq.(\ref{eq:3-loop})
 one finds (see Appendix C) that the perturbative
part  of the condensate remains finite by itself  in the one-loop limit. The same is true for the more 
general
 beta function of 
eq.(\ref{eq:UV-IR-pole}), which is conveniently written  as

\begin{equation}\beta(\alpha)=-\beta_0\alpha^2\ {1+(\omega+{\beta_1\over\beta_0})\alpha+{\lambda\over\beta_0}\alpha^2\over 1+\omega\alpha}
\label{eq:UV-IR-pole-1}\end{equation}
(this is a [2,1] `` Pad\'e  improved'' 4-loop beta function, see  \cite{Steele}).
More precisely, the statement 
 is that, defining $D_{\overline{PT}}(Q^2)$ as in eq.(\ref{eq:PT-bar}) with $\alpha(k^2)$ obtained 
from the solution of the RG equation with the beta function eq.(\ref{eq:UV-IR-pole-1}), the normalization
$\tilde C_{PT} (-1)^{\delta}+C_{PT}$ of the power correction in eq.(\ref{eq:PT-power-corr}) remains finite in the 
limit  $\beta_1, \lambda\rightarrow 0$ with $\lambda/\beta_1$ and $\omega$ fixed (one of these constants may eventually be put
to zero). This limit is taken at fixed ratios $\beta_i/\beta_1$ ($i\geq1$), since one easily checks that 
$\omega=-\beta_3/\beta_2$ and $\lambda/\beta_1=\beta_2/\beta_1-\beta_3/\beta_2$. Note that cancellation occurs
in the one-loop limit
irrespective whether the IR fixed point tends to $+\infty$ (as in the three-loop case) or remains finite (as in the previous
example) .
Similar results hold for
 the [1,2] `` Pad\'e  improved'' 4-loop beta function
\begin{equation}\beta(\alpha)=-\beta_0\alpha^2\ {1+(\tilde{\omega}+{\beta_1\over\beta_0})\alpha\over 1+\tilde{\omega}\alpha+{\tilde{\lambda}\over\beta_0}\alpha^2}
\label{eq:IR-pole-pole}\end{equation}
in the 
limit  $\beta_1, \tilde{\lambda}\rightarrow 0$ with $\tilde{\lambda}/\beta_1$ and $\tilde{\omega}$ fixed.
On the other hand, one finds no cancellation takes place
in the similar  limit $\beta_1, \beta_2\rightarrow 0$
 with $\beta_2/\beta_1$  fixed
if one sets\footnote{No cancellation  probably  occurs either if one sets
 $\omega+{\beta_1\over\beta_0}=0$ in eq.(\ref{eq:UV-IR-pole-1}). I checked only this statement in
 the peculiar case $\omega={\beta_1\over\beta_0}=0$ which corresponds to the 3-loop beta function
with $\beta_1=0$.} $\tilde{\omega}=0$ in eq.(\ref{eq:IR-pole-pole}), which gives

\begin{equation}\beta(\alpha)=-\beta_0\alpha^2\ {1+{\beta_1\over\beta_0}\alpha\over 1-{\beta_2\over\beta_0}\alpha^2}
\label{eq:complex-Landau}\end{equation}
However   in this example higher order beta function coefficients
are further suppressed in the one loop limit compared to $\beta_1$ and $\beta_2$.
 Although the ultimate reason for presence or absence
 of cancellation is not  clear to
 the author,
 the previous examples do suggest that cancellation may be a generic feature, which requires no fine-tuning of parameters
if the one-loop limit is taken at fixed
 and {\em finite} ratios $\beta_{i+1}/\beta_i$ ($i\geq 1$), as follows in QCD from
large $N_f$ counting rules for the class of beta functions which  become one-loop at large $N_f$
(I assume the limit is taken at fixed $\alpha$, which means $\alpha$ contains an
extra factor of $N_f$ compared to the standard\footnote{ With the standard
  normalization of the coupling, the one-loop limit requirement means \cite{Brodsky}
 that $\beta_i$ is at most ${\cal O}(N_f^i)$ for $i\geq 1$.}
normalization).

\section{Conclusions}

1) Landau singularities are usually interpreted as the  signal from the perturbative side of the occurrence
 of non-perturbative phenomena. 
However, in the usual picture at fixed $N_f$, this interpretation is obscured by the fact that they
appear as technical artifacts of  perturbation theory, and would never show up in the full
non-perturbative amplitude. In contrast to this, the previous considerations give
a more direct, {\em physical} motivation for Landau singularities, assuming a two-phase
structure of QCD: they should trace the lower boundary $N_f^*$ of the conformal window. This approach avoids
the notoriously tricky disentangling between the ``perturbative'' and the ``non-perturbative'' parts of the QCD
amplitudes within the conformal window, since they are by definition entirely ``perturbative'' there. On the other
hand, such a separation is naturally achieved below the conformal window, by introducing the analytic continuation
of the conformal window amplitudes to the $N_f<N_f^*$ region. This procedure has been illustrated using the ``APT''
model as a toy ``non-perturbative'' model for the coupling below $N_f^*$. If these ideas are correct, they
reveal a deep connection between information of essentially ``perturbative'' origin (the onset of Landau singularities
on the first sheet) and ``non-perturbative'' phenomena (the emergence of a new phase of QCD when $N_f$ is varied).

When extended to SQCD, these considerations suggest that the NSVZ beta function eq.(\ref{eq:NSVZ}), which has
 a pole at positive
coupling  corresponding to a space-like IR Landau singularity, cannot really be ``non-perturbatively exact'',
 despite it receives
 no contribution from the instanton sector. Instead, the Landau singularity must be present as a signal 
(from the ``perturbative''
side)
 that
 below  Seiberg conformal window \cite{Seiberg} there are other more 
``genuine''
non-perturbative contributions, unrelated to instantons, which remove the singularity 
(the point that instantons do not exhaust
all non-perturbative phenomena is  familiar in QCD). Note the present view differs from the one expressed in \cite{Kogan}.

2) Assuming that the {\em perturbative} QCD coupling has a non-trivial IR fixed point $\alpha_{IR}$
 even in some range {\em below} $N_f^*$ leads
to the equation $\gamma(N_f=N_f^*)=1$ to determine $N_f^*$ from the critical exponent $\gamma$ at the IR fixed point.
Using the available terms in the Banks-Zaks expansion, this equation yields $4\leq N_f^*\leq 6$. It would clearly be desirable to have more terms to better control the
accuracy of the Banks-Zaks expansion.  Note that this condition
is inconsistent with the existence of an analogue of Seiberg duality  in QCD,
 which would rather imply  $\gamma(N_f=N_f^*)=0$.

3) The {\em low} value obtained for $N_f^*$, and its closeness to $N_f=3$, gives some justification to the IR finite
coupling approach to power corrections, and its associated ``universality'' property: the latter could be
violated only by ``genuine non-perturbative terms'' (unrelated
 to the IR finite coupling or to
 an hypothetical non-perturbative modification thereof which may very well not exist)
 which vanish within the conformal window and
 may be expected
to be still small for values such as $N_f=3$ which are  not too far below $N_f^*$. Thus in a first
 approximation 
the bulk of those power corrections  whose very existence is not linked to the non-trivial 
vacuum below $N_f^*$ (condensates which do not break any symmetry of the vacuum, like
 the gluon condensate) should be
 given by the ``perturbative"
 part of the 
condensate contained in the conformal window amplitude $D_{\overline{PT}}(Q^2)$, and related to the
{\em perturbative} IR finite QCD coupling. Consequently, their normalization could 
 even be {\em calculable} from {\em perturbative} input: indeed one obtains $\alpha_{IR}\simeq 0.4$
at the bottom of the conformal window. In this way, the physics of the conformal window
becomes relevant to real-world QCD with $N_f=3$ flavors.

4) Some conditions on the QCD beta function are required: the {\em only} source of  Landau singularities must
arise from the $\alpha>\alpha_{IR}$ region. One needs in
particular
a {\em negative} UV fixed point $\alpha_{UV}$ if perturbation theory is still asymptotic at negative coupling
 for the considered beta function.
 There is indeed some tentative evidence (relying on a modified Banks-Zaks expansion) for such a fixed point in QCD.
 However, the asymptoticity of perturbation theory at negative
coupling, which  points out to a very specific coupling, remains to be understood. An
attractive possibility is to identify this coupling to the ``skeleton coupling'' \cite{Brodsky} associated to a (yet hypothetical) QCD
``skeleton expansion''.

5) A  negative UV fixed point is also required in the SQCD case, 
where duality fixes  the conformal window.

6) The condition $0<\gamma<1$ appears to be  necessary \cite{Gar-Gru-conformal}  for causality under rather
general assumptions. A similar constraint involving the UV fixed point critical exponent  has been derived \cite{Martin-Wells} using completely different arguments as a consistency condition
for non-asymptotically free supersymmetric gauge theories to have a non-trivial physical (positive) UV fixed point.
 
7) It is possible the  IR fixed point  persists in the perturbative QCD coupling down to
 the $N_f\rightarrow -\infty$ one-loop limit. It may even remain {\em finite} in this limit: a simple
example is provided by the beta function eq.(\ref{eq:UV-IR-pole})
 with one positive pole $\alpha_P$ (the required Landau singularity) and two 
opposite sign
 zeroes $\alpha_{IR}$ and $\alpha_{UV}$. The one-loop limit is achieved for $\alpha_{IR}=\alpha_P$
and $\alpha_{UV}=-\infty$. In this example, although $\alpha_{IR}$ remains finite,  $\gamma$ tends to $+\infty$, and therefore necessarily
crosses $1$ and violates causality before the one-loop limit is reached, as expected.

8) The  behavior of conformal window amplitudes (which contain power contributions) has been studied in various
 examples where the IR fixed point is present up to the one-loop limit. The results indicate that, beyond 2-loop, 
finiteness  of the full amplitudes in this limit 
may be a natural and
generic feature requiring no fine tuning (as suggested by
 the behavior of the corresponding perturbative series), provided the limit is taken at fixed and finite ratios of
 the  perturbative
 beta function coefficients beyond one-loop: this is indeed the case for beta functions (such as the one
associated to the ``skeleton coupling'') which become one-loop in
 the $N_f\rightarrow -\infty$ limit of QCD.

\acknowledgments
I thank A. Armoni and A.H. Mueller for useful  discussions.

\appendix

\section{A necessary condition for causality}

Let us show that the condition eq.(\ref{eq:gamma-causal})

\begin{equation}0<\gamma<1\label{eq:gamma-causal1}\end{equation}
is a necessary one for causality. The case where there is an $\alpha>\alpha_{IR}$ Landau singularity was dealt with
in section 5. Let us now consider the alternative situation\footnote{The argument given for this case in \cite{Gar-Gru-conformal} is 
not quite correct.} 
where there is an $\alpha>\alpha_{IR}$ UV fixed point. Then along the ray
eq.(\ref{eq:ray}) there is an ``irrelevant'' real trajectory where the coupling flows from the non-trivial IR fixed point
to the non-trivial UV fixed point. This trajectory is ``irrelevant'' since the correct relevant analytic continuation
of the coupling to the complex momentum plane must always respect the condition of UV asymptotic freedom. This is possible
since the solution of the RG equation along a given ray is not unique, and there can be another (complex)
 solution along the {\em same} ray, which approaches the trivial UV fixed point. The crucial point however is that this alternative
solution cannot approach the non-trivial IR fixed point at low momenta, since the solution along a given ray in the domain of
attraction of a given {\em non-trivial} fixed point is instead  {\em unique} (unless one reaches a Landau singularity),
 as will be shown below. Consequently, the ``relevant''
 solution along the ray eq.(\ref{eq:ray})
which respects UV asymptotic freedom must  in the infrared approach the {\em trivial} $0^-$ IR fixed point. Since
the relevant solution along the space-like axis approaches instead the non-trivial IR fixed point, this means that there must
 be a curve in the complex $k^2$ plane (a  ``separatrix'') which separates the region which is in the domain of attraction
of the non-trivial IR fixed point from the one which is in the domain of attraction
of the trivial IR fixed point. If the condition eq.(\ref{eq:gamma-causal1}) is violated, the ray eq.(\ref{eq:ray}), hence
also the separatrix, belong to the first sheet.
 I shall assume that such an {\em infrared}  separatrix 
(for instance a ray between the space-like axis and the ray
eq.(\ref{eq:ray})) also indicates a discontinuity at finite $k^2$,
 hence the presence of  a Landau singularity on the first sheet along the separatrix. In the 3-loop example of 
eq.(\ref{eq:3-loop}) with $\beta_0>0$, $\beta_2>0$ and $\beta_1<0$, this singularity is the $\alpha<0$ Landau singularity
along the ray eq.(\ref{eq:alpha<0}), whose phase $\pi\vert\beta_1\vert/\beta_0^2$ is indeed intermediate \cite{Gar-Gru-conformal}
in this case between zero (the space-like axis) and the phase
$\pi/\gamma$ of the ray eq.(\ref{eq:ray}).

To prove unicity of the solution of the renormalization group equation  along a given ray, under the constraint this solution
 approaches
a given {\em non-trivial} IR fixed point in the infrared, consider the solution eq.(\ref{eq:alpha-low-low}) of this
equation
 around $\alpha=\alpha_{IR}$. Inverting to solve for $\alpha_{IR}-\alpha$ yields the {\em unique} power series solution
(with $(k^2/\Lambda^2)^\gamma$ as expansion parameter)

\begin{equation}\alpha_{IR}-\alpha=\left({k^2\over \Lambda^2}\right)^{\gamma}-
{\gamma_1\over \gamma}\left({k^2\over \Lambda^2}\right)^{2{\gamma}}+...
\label{eq:IRexpansion}
\end{equation}
which suggests that there is a unique solution, at least in a neighborhood of $k^2=0$.

Note the corresponding unicity for the expansion around a {\em trivial} fixed point does not hold. Consider  the
weak coupling solution eq.(\ref{eq:alpha-2loop}) of the RG equation. One finds that
there are  a priori {\em two} possible  solutions which approach the trivial IR fixed point $0^-$, i.e. such that
 $\vert\alpha(k^2)\vert
\rightarrow 0$ for $k^2\rightarrow 0$, along the rays $k^2=\vert k^2\vert \exp(\pm i\pi\beta_1/\beta_0^2)$. Putting
$\alpha\equiv \vert\alpha\vert \exp(i\phi)$, one solution corresponds to $\alpha<0$, i.e. $\phi\equiv \pm\pi$, while the other
gives {\em complex} values of $\alpha$, i.e. $\phi\equiv \phi(k^2)\neq 0,\pm\pi$ with $\phi(k^2) \rightarrow \pm\pi$
(and $\sin(\phi)/\vert\alpha\vert \rightarrow\pm 2\pi\beta_1/\beta_0$)
for
$k^2\rightarrow 0$. Indeed the real part
of eq.(\ref{eq:alpha-2loop}) yields

\begin{equation}\log\vert {k^2\over \Lambda^2}\vert={1\over\beta_0\vert\alpha\vert}\cos\phi+...\label{eq:alpha-2loop-real}
\end{equation}
while the imaginary part yields

\begin{equation}\pm\pi{\beta_1\over\beta_0^2}=-{1\over\beta_0\vert\alpha\vert}\sin\phi+{\beta_1\over\beta_0^2}\phi+...
\label{eq:alpha-2loop-imaginary}
\end{equation}
from which the previous results follow.
Eventually we also have $\phi(k^2) \rightarrow 0$ for $k^2\rightarrow \infty$ so that $\alpha(k^2)$ approaches
the trivial  UV fixed point $0^+$ if this possible
 ``relevant''
solution is actually realized, as  in the standard two-loop case with $\beta_1/\beta_0 >0$ 
(where the $\alpha<0$ solution is attracted
towards the non-trivial (negative) UV fixed point, and is therefore ``irrelevant''). A similar situation arises in the 3-loop
case of eq.(\ref{eq:3-loop}) with $\beta_0>0$, $\beta_2<0$ and $\beta_1>0$ and large enough so that the phase
$\pi/\gamma$ of the ray eq.(\ref{eq:ray}) is intermediate between zero and the phase $\pi\beta_1/\beta_0^2$ of the ray
eq.(\ref{eq:alpha<0}).

\section{A modified Banks-Zaks expansion}

One introduces the auxiliary beta function

\begin{equation}{\tilde \beta}(\alpha)\equiv -\beta_0 \alpha^2-{\tilde\beta_1} \alpha^3-\beta_2 \alpha^4-\beta_3 \alpha^5+...
\label{eq:beta-tilde}\end{equation}
where the two-loop coefficient is replaced by ${\tilde\beta_1}\equiv \rho\beta_0$ where $\rho$ is an arbitrary constant, and 
solves perturbatively the equation ${\tilde \beta}(\alpha)=0$ for $\beta_0\rightarrow 0$, i.e. $N_f\rightarrow 16.5$
at {\em fixed} $\rho$.
Then for each value of $N_f$, one adjusts   $\rho$ in the resulting series to the value $\rho=\beta_1/\beta_0$ for which
${\tilde \beta}(\alpha)$ coincides with the true beta function $\beta(\alpha)$. This procedure yields the modified Banks-Zaks
series for a couple of (IR,UV) fixed points

\begin{equation}{\tilde \alpha}_{IR,UV}=\pm {\tilde \epsilon}+ {1\over 2}\left(\rho- {\beta_{3,0}\over \beta_{2,0}}\right) {\tilde \epsilon}^2+...
\label{eq:BZ-modified}\end{equation}
with ${\tilde \epsilon}\equiv (-\beta_0/\beta_{2,0})^{1/2}$. For the auxiliary critical exponent

\begin{equation}{\tilde \gamma}\equiv
d{\tilde\beta}(\alpha)/ d\alpha\vert_{\alpha={\tilde\alpha_{IR}}}\label{eq:exponent-aux}\end{equation}
one gets the expansion

\begin{equation}{\tilde \gamma}= 2\ \beta_0\ {\tilde \epsilon}(1+\rho{\tilde \epsilon}+...)\label{eq:BZ-modified-exponent}
\end{equation}
For  $\rho=0$ (which corresponds to $N_f=8.05$) one recovers the results of section 7, while for $\rho=-2.55$ 
(which corresponds to $N_f=11$) one gets a next-to leading order correction of about $80\%$ in eq.(\ref{eq:BZ-modified}) and of $60\%$
in eq.(\ref{eq:BZ-modified-exponent}) if one uses the pinch technique ``skeleton coupling''. 
Going to even larger $N_f$ values will presumably not help: although the next-to-leading order correction in the fixed point
series  decreases (due to a cancellation between  $\rho$ and $-\beta_{3,0}/ \beta_{2,0}$), it  increases in the critical exponent series due to the larger
 value of $\rho$, conveying the
suspicion that still higher order corrections in eq.(\ref{eq:BZ-modified}) will be large (as they must since one does not
expect the negative UV fixed point, at the difference of the IR fixed point, to approach zero as $\beta_0\rightarrow 0$).

\section{Power corrections from IR finite perturbative coupling}

 I first recall the results of \cite{Grunberg-FP,D-U} to compute the power correction in eq.(\ref{eq:PT-power-corr}) when the
perturbative coupling $\alpha(k^2)$ in eq.(\ref{eq:PT-bar}) has a non-trivial IR fixed point. The main observation
is that $D_{\overline{PT}}(Q^2)$ remains finite (and approaches $\alpha_{IR}$) for $Q^2\rightarrow 0$.
 Eq.(\ref{eq:PT-power-corr}) then implies that   in the same limit  the Borel sum eq.(\ref{eq:Borel-sum}) diverges as
\begin{equation}D_{PT}(Q^2) \sim -[\tilde C_{PT} (-1)^{\delta}+C_{PT}]\left({\Lambda^2\over Q^2}\right)^n
\label{eq:PT-IR}
\end{equation}
 This behavior can be reproduced with the following ansatz for  the Borel image
 $\tilde D(z)$ of $D_{PT}(Q^2)$ . Put 

\begin{equation}\tilde D(z)\equiv \exp\left ({z\over \alpha_{IR}}\right) F(z)\label{eq:Borel-ansatz}
\end{equation}
and assume 

\begin{equation}F(z)\sim {\kappa\over z^a}\label{eq:Borel-large}
\end{equation}
as $z \rightarrow +\infty$ (with $a<1$). Then eq.(\ref{eq:Borel-sum}) implies, since the integral is dominated by the large
$z$ contribution for $Q^2\rightarrow 0$

\begin{equation}D_{PT}(Q^2) \sim \kappa \int_0^\infty dz \exp\left(-z/A(Q^2)\right) {1\over z^a}=\kappa\ 
\Gamma(1-a) A(Q^2)^{1-a}
\label{eq:Borel-IR1}\end{equation}
where

\begin{equation}{1\over A(Q^2)}\equiv {1\over \alpha(Q^2)}-{1\over \alpha_{IR}}\label{eq:A-coupling}\end{equation}
 In the IR limit we get from
 eq.(\ref{eq:alpha-low}) that
 
\begin{equation}A(Q^2)\sim  \left({\Lambda^2\over Q^2}\right)^{\gamma}\label{eq:A-coupling-low}\end{equation} 
(the normalization implies a proper redefinition
of $\Lambda$), hence

\begin{equation}D_{PT}(Q^2) \sim \kappa\ \Gamma(1-a) \left({\Lambda^2\over Q^2}\right)^{\gamma(1-a)}\label{eq:Borel-IR2}
\end{equation}
Comparing with eq.(\ref{eq:PT-IR}), eq.(\ref{eq:Borel-IR2}) determines 

\begin{equation}\tilde C_{PT} (-1)^{\delta}+C_{PT}=-\kappa\ \Gamma(1-a)\label{eq:Borel-IR3}
\end{equation}
from the  $z\rightarrow +\infty$ behavior of $\tilde D(z)$, and reveals that 

\begin{equation}n=\gamma\ (1-a)\label{eq:a-parameter}
\end{equation}
 To determine the normalization constant $\kappa$  in eq.(\ref{eq:Borel-large}),
one uses the fact that $D_{\overline{PT}}(Q^2)$ as well as $D_{PT}(Q^2)$ satisfy the differential equation \cite{D-U}

\begin{equation}D(\alpha)+{1\over n}\beta(\alpha){dD\over d\alpha}=\alpha\label{eq:equadiff}
\end{equation}
where $\alpha\equiv \alpha(Q^2)$, which yields an integral equation for $\tilde D(z)$

\begin{equation}\int_0^\infty dz \exp\left(-z/\alpha\right)\tilde D(z)+{1\over n}{\beta(\alpha)\over \alpha^2}
\int_0^\infty dz \exp\left(-z/\alpha\right)
 z\tilde D(z)=\alpha\label{eq:equaBorel}
\end{equation}
Introducing the Borel representation of ${\beta(\alpha)\over \alpha^2}$, one can get an equation \cite{Grunberg-FP}
 involving only Borel space
functions. 

Let us apply this method to  the [2,1] ``Pad\'e  improved''  4-loop
 beta function\footnote{The 
[1,2] ``Pad\'e  improved''  4-loop beta function eq.(\ref{eq:IR-pole-pole}) can be dealt with similar methods (one obtains 
an inhomogeneous
 differential equation instead of eq.(\ref{eq:equadiffBorel-ex})), and leads to analogous results.}
$\beta(\alpha)=-\beta_0\alpha^2\ {1+(\omega+{\beta_1\over\beta_0})\alpha+{\lambda\over\beta_0}\alpha^2\over 1+\omega\alpha}$ 
of eq.(\ref{eq:UV-IR-pole-1}).
Multiplying both sides of eq.(\ref{eq:equaBorel}) by the beta function denominator $1+\omega\alpha$, one gets 

\begin{eqnarray}& &(1+\omega\alpha)\int_0^\infty dz \exp\left(-z/\alpha\right)\tilde D(z)-{1\over z_n}
\left[1+\left(\omega+{\beta_1\over\beta_0}\right)\alpha+{\lambda\over\beta_0}\alpha^2\right]
\int_0^\infty dz \exp\left(-z/\alpha\right)
 z\tilde D(z)\nonumber\\
& &=\alpha(1+\omega\alpha)\label{eq:equaBorel-ex}\end{eqnarray}
where $z_n={n\over\beta_0}$. Using the convolution multiplication theorem for Borel transforms eq.(\ref{eq:equaBorel-ex}) becomes in Borel space

\begin{eqnarray}& &\tilde D(z)+\omega \int_0^z dy \tilde D(y)-{1\over z_n}\left[z\tilde D(z)+\left(\omega+{\beta_1\over\beta_0}\right)
\int_0^z dy\ y \tilde D(y)+{\lambda\over\beta_0} \int_0^z dy (z-y)y \tilde D(y)\right]\nonumber\\
& &=1+\omega z\label{eq:equaBorel1-ex}\end{eqnarray}
After taking two derivatives with respect to $z$, eq.(\ref{eq:equaBorel1-ex}) yields the homogenous second order differential equation
with linear coefficients

\begin{equation}\left(1-{z\over z_n}\right)\tilde D^{''}(z)-\left[{2\over z_n}-\omega+\left(\omega+{\beta_1\over\beta_0}\right){z\over z_n}\right]
\tilde D^{'}(z)-\left[{1\over z_n}\left(\omega+{\beta_1\over\beta_0}\right)+{\lambda\over\beta_0}{z\over z_n}\right]\tilde D(z)=0
\label{eq:equadiffBorel-ex}\end{equation}
which is equivalent to eq.(\ref{eq:equaBorel-ex}) with the boundary conditions

\begin{eqnarray}\tilde D(0)&=&1\nonumber\\
& &\label{eq:bound-cond}\\
\tilde D^{'}(0)&=&{1\over z_n}\nonumber\end{eqnarray}
This equation can be solved in a standard way \cite{Bateman} in term of confluent hypergeometric functions. If one writes
$\tilde D(z)$ as in eq.(\ref{eq:Borel-ansatz}), then $F(z)\equiv F(a,c;x)$,
 a solution of the confluent hypergeometric
equation

\begin{equation}x{d^2F\over dx^2}+(c-x){dF\over dx}-aF=0\label{eq:hypergeo}\end{equation}
where

\begin{equation}x=x_0\left(1-{z\over z_n}\right)\label{eq:variable-x}\end{equation}
with 
 
 \begin{equation}x_0=({1\over \alpha_{IR}}-{1\over \alpha_{UV}})z_n\label{eq:x0}\end{equation}
and $\alpha_{IR}$ and $\alpha_{UV}$ are
the positive and negative zeroes of the beta function eq.(\ref{eq:UV-IR-pole-1}). Moreover we have

\begin{equation}c=2+{\beta_1\over \beta_0}z_n\equiv 2+\delta\label{eq:param-c}\end{equation}
whereas the parameter $a$ is related to the IR fixed point critical exponent $\gamma$ by eq.(\ref{eq:a-parameter}). In 
term of $F(a,c;x)$, the boundary conditions eq.(\ref{eq:bound-cond}) read

\begin{eqnarray}F(a,c;x_0)&=&1\nonumber\\
& &\label{eq:bound-cond1}\\
x_0{dF\over dx}\vert_{x=x_0}& =&{z_n\over \alpha_{IR}} - 1\nonumber\end{eqnarray}
The general solution of eq.(\ref{eq:hypergeo}) is 

\begin{equation}F(a,c;x)=F_{sing}(a,c;x)+F_{reg}(a,c;x)\label{eq:sol}\end{equation}
where in standard notation \cite{Bateman}

\begin{equation}F_{reg}(a,c;x)=K_r\ \Phi(a,c;x)\equiv K_r\ \Phi(a,2+\delta;x)\label{eq:F-reg}\end{equation}
and

\begin{equation}F_{sing}(a,c;x)=K_s\ x^{1-c}\ \Phi(a-c+1,2-c;x)\equiv K_s\ \left({1\over x}\right)^{1+\delta}\ 
\Phi(a-1-\delta,-\delta;x)\label{eq:F-sing}\end{equation}
where $K_r$ and $K_s$ are integration constants to be determined from eq.(\ref{eq:bound-cond1}).
To find $\kappa$ in eq.(\ref{eq:Borel-large}), one needs the $z\rightarrow +\infty$ behavior, which corresponds
 (eq.(\ref{eq:variable-x})) to $x\rightarrow -\infty$, since $x_0>0$. One gets \cite{Bateman}
 in this limit 

\begin{equation}F_{sing}(a,c;x)\sim \kappa_s\  {1\over z^a}\label{eq:Fas-sing}\end{equation}
with

\begin{equation}\kappa_s=K_s\ {\Gamma(-\delta)\over \Gamma(1-a)}(-1)^{1+\delta}\left({z_n\over x_0}\right)^a\label{eq:kap-s}
\end{equation}
and

\begin{equation}F_{reg}(a,c;x)\sim \kappa_r\  {1\over z^a}\label{eq:Fas-reg}\end{equation}
with

\begin{equation}\kappa_r=K_r\ {\Gamma(2+\delta)\over \Gamma(2+\delta-a)}\left({z_n\over x_0}\right)^a\label{eq:kap-s}
\end{equation}
Note that $\kappa_s$ is complex, corresponding to the ``renormalon contribution''. Since

\begin{equation}\kappa=\kappa_r+\kappa_s\label{eq:kap}\end{equation}
one obtains
comparing with eq.(\ref{eq:Borel-IR3})

 \begin{equation}\tilde C_{PT}=K_s\left({z_n\over x_0}\right)^a\Gamma(-\delta)\label{eq:C-s}
\end{equation}
and

\begin{equation} C_{PT}=-K_r\left({z_n\over x_0}\right)^a{\Gamma(2+\delta)\over\Gamma(2+\delta-a)}\Gamma(1-a)\label{eq:C-r}
\end{equation}
Thus the normalization of the ``condensate''  in this model is

\begin{equation}\tilde C_{PT} (-1)^{\delta}+C_{PT}=\left({z_n\over x_0}\right)^a \left[(-1)^{\delta}K_s\ \Gamma(-\delta)-
K_r\ {\Gamma(2+\delta)\over\Gamma(2+\delta-a)}\Gamma(1-a)\right]\label{eq:condensate}
\end{equation}
with $K_r,K_s$ determined by eq.(\ref{eq:bound-cond1}) as mentioned above. Using the relation
${d\Phi(a,c;x)\over dx}={a\over c}\Phi(a+1,c+1;x)$,
this equation yields 

\begin{eqnarray}K_s \left({1\over x_0}\right)^{1+\delta}\Phi(a-1-\delta,-\delta;x_0)+K_r\ \Phi(a,2+\delta;x_0)&=&1\nonumber\\
& &\label{eq:bound-cond2}\\
K_s\left[(-1-\delta)\left({1\over x_0}\right)^{1+\delta}\Phi(a-1-\delta,-\delta;x_0)+\left({1\over x_0}\right)^{\delta}
\left({a-1-\delta\over -\delta}\right)\Phi(a-\delta,1-\delta;x_0)\right]&+& \nonumber\\
K_r\ {a\over 2+\delta}\ x_0\ \Phi(a+1,3+\delta;x_0)&=&
{z_n\over \alpha_{IR}}-1\nonumber\end{eqnarray}

Consider now the one-loop limit where $\beta_1$ and $\lambda \rightarrow 0$ with $\omega$ and $\lambda/\beta_1$ fixed.
This limit corresponds to fixed ratios $\beta_i/\beta_1$ ($i>1$), for instance we have

\begin{eqnarray}{\beta_2\over \beta_0}&=&-\omega (1-r){\beta_1\over \beta_0}\nonumber\\
& &\label{eq:ratios}\\
{\beta_3\over \beta_0}&=&\omega^2 (1-r){\beta_1\over \beta_0}\nonumber\end{eqnarray}
where $r\equiv {\lambda\over \beta_1}{1\over\omega}$ (note that $\omega=-\beta_3/\beta_2$).
In 
this limit  $\gamma\rightarrow +\infty$, hence (eq.(\ref{eq:a-parameter})) $a\rightarrow 1$, $\delta\rightarrow 0$,
while $x_0\sim{z_n\over \alpha_{IR}}\sim -z_n\omega$  remains finite (I assume presently $\omega<0$ and $\neq 0$ to have the
beta function pole positive, finite).
 Moreover a little algebra shows that
${a-1-\delta\over -\delta}\sim -{\lambda\over \beta_1}\alpha_{IR}\sim r$ is finite. This implies, using the
 small x expansion of $\Phi(a,c;x)$, that $\Phi(a-1-\delta,-\delta;x_0)\sim
1+r(e^{x_0}-1)$.
 Using also the  special values \cite{Bateman} of the $\Phi$ function,
 the system eq.(\ref{eq:bound-cond2}) then simplifies to

\begin{eqnarray}K_s\ {1\over x_0}\left[1+r(e^{x_0}-1)\right]-K_r\ {1\over x_0}(1-e^{x_0})&\sim&1\nonumber\\
& &\label{eq:bound-cond3}\\
K_s\left[-{1\over x_0}\left(1+r(e^{x_0}-1)\right)+re^{x_0}\right]+K_r\ {1\over x_0}\left[1-e^{x_0}(1-x_0)\right]&\sim&x_0-1
\nonumber\end{eqnarray}
which gives the one-loop limit values

\begin{eqnarray}K_s&\sim& x_0e^{-x_0}\nonumber\\
& &\label{eq:K-values}\\
K_r&\sim&(1-r)x_0e^{-x_0}\nonumber\end{eqnarray}
Finally in this limit eq.(\ref{eq:condensate}) yields

\begin{eqnarray}\tilde C_{PT} (-1)^{\delta}+C_{PT}&\sim & \left({z_n\over x_0}\right)^a\left(-{K_s\over\delta}-{K_r\over 1-a}\right)
\nonumber\\
& &\label{eq:condensate-limit}\\
&\sim &-z_n\ e^{-x_0}\left({1\over\delta}+{1-r\over 1-a}\right)\nonumber\end{eqnarray}
where  the two separately divergent 
 terms 
 in eq.(\ref{eq:condensate-limit})  cancel in leading order 
as a consequence of the previously mentioned relation ${a-1-\delta\over -\delta}\sim r$, leaving a {\em finite}
 one-loop limit. Note that $\tilde C_{PT}$ diverges as

\begin{equation}\tilde C_{PT}\sim -{\beta_0\over \beta_1}e^{-x_0}\label{eq:condensate-limit1}\end{equation}
in agreement with the general expectation eq.(\ref{eq:C-sing}).
Note also we should have $r>1$ if $\beta_1>0$ or $r<0$ if $\beta_1<0$ in order that $\alpha_{UV}<0<\alpha_{IR}<\alpha_{P}$
(the case $0<r<1$ is also  possible if $\beta_1<0$, and corresponds to a situation where 
$\alpha_{P}<\alpha_{UV}<0<\alpha_{IR}$).

The case where $\omega=0$ ($r=\infty$) corresponds to the standard
 three-loop beta function eq.(\ref{eq:3-loop}), and has to be treated separately.
 Here too, one finds that cancellation occurs generically, provided the one loop limit is taken at fixed and {\em finite} ratio
$\beta_2/\beta_1$ (there is no cancellation if $\beta_1 \equiv 0$). Then a little algebra shows that $a\rightarrow 1$ and $\delta\rightarrow 0$ as before,
while this time $x_0\sim 2{z_n\over \alpha_{IR}}\rightarrow 0$. Moreover one finds that
$\left({a-1-\delta\over -\delta}\right)x_0\sim -{\beta_2\over \beta_1}z_n$ is finite. The system eq.(\ref{eq:bound-cond2}) 
simplifies in this limit to

\begin{eqnarray}K_s\ {1\over x_0}\left(1-{\beta_2\over \beta_1}z_n\right) +K_r&\sim&1\nonumber\\
& &\label{eq:bound-cond3}\\
K_s\left[-{1\over x_0}\left(1-{\beta_2\over \beta_1}z_n\right)+ {\beta_0\over 2\beta_1 \alpha_{IR}}\right]+
{1\over 2}x_0\ K_r&\sim&
-1\nonumber\end{eqnarray} 
which gives, using $x_0\sim 2{z_n\over \alpha_{IR}}\rightarrow 0$ and $\alpha_{IR}^2\sim -{\beta_0\over \beta_2}$ 
(one must
  take $\beta_2<0$ to have an IR fixed point present down to the one-loop limit)

\begin{eqnarray}K_s\ {1\over x_0}\left(1-{\beta_2\over \beta_1}z_n\right) +K_r&\sim&1\nonumber\\
& &\label{eq:bound-cond4}\\
-K_s\ {1\over x_0}+{1\over 2}x_0\ K_r&\sim&-1\nonumber\end{eqnarray}
and yields

\begin{eqnarray}K_s&\sim& x_0\rightarrow 0\nonumber\\
& &\label{eq:bound-cond5}\\
K_r&\sim& {\beta_2\over \beta_1}z_n\nonumber\end{eqnarray}
From eq.(\ref{eq:C-s}) and  (\ref{eq:C-r}) one then gets after a little algebra the  values

\begin{equation}\tilde C_{PT}\sim -{\beta_0\over\beta_1}\label{eq:condensate-limit1}\end{equation}
and

\begin{equation} C_{PT}\sim {\beta_0\over\beta_1}\label{eq:condensate-limit2}\end{equation}
which again do cancel in leading order, leaving a {\em finite} one-loop limit condensate.


\end{document}